\documentclass[12pt]{article}
\hoffset = - 0.5 truecm \voffset=-1.4 truecm
\def\beq{\begin{equation}}   \def\eeq{\end{equation}}
\def\bea{\begin{eqnarray}}  \def\eea{\end{eqnarray}} \def\nn{\nonumber}
\def\noi{\noindent} \def\beeq{\begin{eqnarray}}
\def\eeeq{\end{eqnarray}}
\def\lsim{\raise0.3ex\hbox{$<$\kern-0.75em\raise-1.1ex\hbox{$\sim$}}}
\def\gsim{\raise0.3ex\hbox{$>$\kern-0.75em\raise-1.1ex\hbox{$\sim$}}}

\usepackage{graphicx}
\usepackage{epsfig}
\newcommand\mysection{\setcounter{equation}{0}\section}
\renewcommand{\theequation}{\thesection.\arabic{equation}}
\newcounter{hran} \renewcommand{\thehran}{\thesection.\arabic{hran}}

\def\bmini{\setcounter{hran}{\value{equation}}
          \refstepcounter{hran}\setcounter{equation}{0}
          \renewcommand{\theequation}{\thehran\alph{equation}}\begin{eqnarray}}

\def\bminiG#1{\setcounter{hran}{\value{equation}}
\refstepcounter{hran}\setcounter{equation}{-1}
\renewcommand{\theequation}{\thehran\alph{equation}}
\refstepcounter{equation}\label{#1}\begin{eqnarray}}

\def\emini{\end{eqnarray}\relax\setcounter{equation}{\value{hran}}\renewcommand{\theequation}{\thesection.\arabic{equation}}}

\begin{document}

\begin{titlepage}
 
\begin{flushright}

LAPTH-1091/05\\
LPT-Orsay 04-53\\

\end{flushright}
\vspace{1.cm}

\begin{center}
\vbox to 1 truecm {}
{\large \bf  New NLO Parametrizations of the Parton Distributions in
Real Photons}

\vskip 1 truecm
{\bf P. Aurenche$^{(a)}$, M. Fontannaz$^{(b)}$, J. Ph.
Guillet$^{(a)}$} \vskip 3 truemm

{\it $^{(a)}$ LAPTH, UMR 5108 du CNRS associ\'ee \`a l'Universit\'e
de Savoie, \\ BP 110, Chemin de Bellevue, 74941 Annecy-le-Vieux
Cedex, France}

\vskip 3 truemm

{\it $^{(b)}$ Laboratoire de Physique Th\'eorique, UMR 8627 CNRS,\\
Universit\'e Paris XI, B\^atiment 210, 91405 Orsay Cedex, France}
\vskip 2 truecm

\begin{abstract}
We present new NLO sets of parton distributions in real photons based
on a scheme invariant definition of the non-perturbative input. We
compare the theoretical predictions with LEP data and a best fit allows
us to constrain the parameters of the distributions. The shape of the
gluon distribution is poorly constrained and we consider the possibility
to measure it in photoproduction experiments. Three parametrizations
which aim to take into account the scattering of LEP data are
proposed. They are compared to other NLO parametrizations.
\end{abstract}
\end{center}
\end{titlepage}

\pagestyle{plain}
\baselineskip=22 pt
\mysection{Introduction}
\hspace*{\parindent}
Since the early days of QCD, the photon structure function has attracted
much interest, and the pioneering work of Witten \cite{2r} triggered a
large amount of theoretical and experimental studies \cite{1r}. Recent
developments are nicely reviewed in ref. \cite{3r,4r}. The present
situation is characterized by much recent data, essentially accumulated
by LEP experiments, by the possibility to observe the photon structure
function in photoproduction experiments at HERA \cite{5r}, and by the
necessity to have accurate predictions for the Next Linear Collider
(NLC). These three reasons justify an upgrading of the AFG
parametrization of quark and gluon distributions in the photon that we
proposed ten years ago \cite{6r}. \par

The NLO AFG parametrization was characterized by a non-perturbative
input, defined in a factorization-scheme invariant way, and by a parameter
$Q_0^2$ fixing the starting point of the $Q^2$-evolution of the
perturbative component. With the choice $Q_0^2 = .5$~GeV$^2$ (a value
close to the $\rho$-mass squared) and a non-perturbative input determined
within the framework of the Vector Dominance Model (VDM), we found
good agreement with data.  \par

Data on the photon structure function essentially determine
the quark content of the photon. On the other
hand the gluon content can be constrained in photoprod\-uction reactions
at HERA \cite{5r,9r} and the AFG gluon distribution appears to be in
agreement with recent data on jet production \cite{10r}. However the latter
lacks flexibility and a parametrization containing adjustable parameters
should allow a better fit of the relevant data. In particular the VDM
input used in the AFG parametrization rests on the $\pi^0$ structure
function determined from prompt photon and Drell-Yan experiments
\cite{11r} and the user is not allowed to modify this input. Moreover,
the parametrization was only valid for $N_f = 4$~; the large energies
reached in collider experiments now require that we take into account the
bottom quark contribution.\par

The new AFG04 parametrization of the quark and gluon distributions in
the real photon is valid for $N_f = 5$. We work at the NLO
approximation and within the massless, flavor changing scheme. However
we keep $m_q^2/Q^2$ corrections ($q = c,b$) in the direct contribution
in order to have smooth thresholds when calculating
$F_2^{\gamma}(x,Q^2)$. Asymptotically, when $m_q^2/Q^2$ goes to zero,
we recover the usual $\overline{MS}$ factorization scheme for massless
partons. The non-perturbative input, always inspired by the VDM
approximation, has a flexible parametrization~: the gluon and the sea
normalization, as well as the gluon shape can be modified. The overall
normalization of the non-perturbative input is also left free, and the
perturbative parameter $Q_0^2$ can be varied. We study the
effects on $F_2^{\gamma}$ of the variation of these
parameters~; constraints are obtained from the confrontation of the
theoretical predictions with LEP data.  As expected data on
$F_2^{\gamma}$ do not give access to the gluon content of the photon.
A better determination of the latter should be obtained from
large-$p_{\bot}$ photoproduction reactions that we briefly
consider. A default parametrization results from these studies. Other
parametrizations, which reflect the scattering of
LEP data, are also proposed.

In section 2, we discuss the necessity to introduce a
scheme-independent non-perturbative input. The method to reach this
goal is detailed in section 3 and appendix A. In section 4, we present
a specific non-perturbative input obtained from the Vector Dominance
Model. Section 5 is devoted to the study of medium-$Q^2$ LEP data,
which allows us to constrain the parametrization of the distributions.
We propose three different distributions which take into account the
scattering of LEP data, and compare our best fit parametrization to the
GRS \cite{8r} and CJK \cite{reference11} NLO parametrizations. Finally
the gluon distribution is considered in detail in section 6. Appendix A
presents a derivation of the scheme-invariant non-perturbative
formalism, and Appendix B presents the parametrizations of the parton
distributions available in the form of a FORTRAN code.

\mysection{Scheme Invariant Non-Perturbative Input}
\hspace*{\parindent}
In this section we recall the method we used \cite{6r} to study the link
between the non-perturbative and the perturbative components of the
photon structure function. Once this link is understood, a
factorization scheme invariant non-perturbative component can be
defined. \par

Let us start with a few definitions. The evolutions of the gluon distribution
$G^{\gamma}(x, Q^2)$, of the singlet distribution $\Sigma^{\gamma}(x,
Q^2) = \sum\limits_{f=1}^{N_f} (q_f^{\gamma} (x, Q^2) + \bar{q}_f^{\gamma} (x,
Q^2)) \equiv \sum_f q_f^{(+)}(x, Q^2)$ and of the non-singlet
distributions $q_f^{NS}(x,Q^2) = q_f^{(+)} - \Sigma^{\gamma}/N_f$ ($N_f$
is the number of flavors) are governed by the inhomogeneous DGLAP
equations \cite{new10}

\bminiG{2.1e}
\label{2.1ae}
{\partial \Sigma^{\gamma} \over \partial \log Q^2} = k_q + P_{qq}
\otimes \Sigma^{\gamma} + P_{qg} \otimes G^{\gamma} \eeeq \beeq
         \label{2.1be}
{\partial G^{\gamma} \over \partial \log Q^2} = k_g + P_{gq}
\otimes \Sigma^{\gamma} + P_{gg} \otimes G^{\gamma}
         \emini

\beq
\label{2.2e}
{\partial q_f^{NS} \over \partial \log Q^2} = \sigma_f^{NS} k_q + P_{NS}
\otimes q_f^{NS}
\eeq

\noi where $\sigma_f^{NS} = (e_f^2/<e^2>-1)/N_f$ with $<e^m> =
\sum\limits_f e_f^m/N_f$. The convolution $\otimes$ is defined by

\beq
\label{2.3e}
P \otimes q = \int_x^1 {dz \over z} \ P \left ( {x \over z}\right ) q(z) \ .
\eeq

The homogeneous $(P_{ij})$ splitting
functions were calculated in ref. \cite{new11,new12} at the NLO
approximation. The inhomogeneous splitting functions

\beq \label{2.4e} k_q = {\alpha \over 2 \pi} \ k_q^{(0)} + {\alpha
\over 2 \pi}\ {\alpha_s (Q^2) \over 2 \pi}\ k_q^{(1)} \eeq

\beq \label{2.5e} k_g = {\alpha \over 2 \pi} \ {\alpha_s (Q^2) \over 2
\pi}\ k_g^{(1)} \eeq

\noindent may be derived from the $P_{ij}$ and are given in refs.
\cite{12r,13r}~; the expression for the LO splitting function
$k_q^{(0)}$ is $2N_f<e^2>\ 3[x^2 + (1 - x)^2]$\footnote{We do not
consider NNLO corrections. Therefore our parametrizations are
consistent with the NLO calculations of large-$p_T$ photoproduction
cross sections {\protect\cite{5r,new15}}. Expressions of NNLO corrections and
discussions of their importance may be found in
{\protect\cite{new16,new17,new18}}.}. \par

In terms of the parton distributions, the photon structure function is
written

\beq \label{2.7e} {\cal F}_2^{\gamma}(x,Q^2) \equiv F_2^{\gamma}(x,
Q^2)/x = \sum_f e_f^2 \ q_f^{(+)} \otimes C_q + G^{\gamma}
\otimes C_g + C_{\gamma} \ . \eeq

\noi The Wilson coefficients $C_q$ and $C_g$ may be found in ref.
\cite{14r}, and the direct term $C_{\gamma}$, in the $\overline{MS}$
scheme, is given by \cite{new.a,14r}

\beq \label{2.8e}
C_{\gamma} = {\alpha \over 2 \pi}\ 2 \sum_f e_f^4 \ 3 \left [ \left
(x^2 + (1 - x)^2 \right ) \ln \ {1 - x \over x} + 8x (1 - x) - 1
\right ] \ . \eeq

The physical quantity ${\cal F}_2^{\gamma}$ is factorization scheme
independent. This means that it does not depend on the procedure (the
factorization scheme) used to define the NLO splitting function
$P_{ij}^{(n)}$ ($n \geq 1$) and $k_i^{(n)}$, and the function $C_q$,
$C_g$ and $C_{\gamma}$. This is however true only if these functions
were calculated to all orders in $\alpha_s$. If the truncated series
(\ref{2.4e}) and (\ref{2.5e}) are used, the photon structure function
is still scheme independent, but only at order ${\cal O}(\alpha_s^0)$.\par

Let us consider, for the sake of simplicity, the non-singlet eq.
(\ref{2.2e}). Its
solution can be written, for moments of the quark distribution
$q_f^{NS}(n) = \int_0^1 dx x^{n-1} q_f^{NS}(x,Q^2)$, as follows~:

\beq \label{2.9e} q_f^{NS}(n) = \sigma_f^{NS} \int^{Q^2} {dk^2 \over k^2}
k_q(n) e^{\int_{k^2}^{Q^2} {dk'^2 \over k'^{2}} P_{NS}(n)} \ . \eeq

\noi For small values of $k^2$, the perturbative approach is no longer
valid. Let us assume that we can use this expression for $k^2 \geq Q_0^2$~;
we then define the perturbative (Anomalous \cite{2r}) component

\beq \label{2.10e} q_{AN}^{NS}(Q^2,Q_0^2) = \sigma^{NS}
\int_{Q_0^2}^{Q^2} {dk^2 \over k^2}
k_q e^{\int_{k^2}^{Q^2} {dk'^2 \over k'^2} P_{NS}}
\eeq

\noi (we have dropped the indices $f$ and $n$). \par

For $k^2$ smaller than $Q_0^2$, we are in the realm of non-perturbative
QCD and we write the corresponding hadronic contribution (which behaves
like a hadron structure function and is discussed in detail in
appendix A)

\beq \label{2.11e} q_{H}^{NS}(Q^2,Q_0^2) = q_{H}^{NS}(Q_0^2) \
e^{\int_{Q_0^2}^{Q^2} {dk'^2 \over k'^{2}} P_{NS}} \ , \eeq

\noi the total non-singlet distribution being the sum of the Anomalous
and Hadronic component

\beq \label{2.12e} q^{NS}(Q^2) = q_{AN}^{NS}(Q^2, Q_0^2) +
q_{H}^{NS}(Q^2,Q_0^2) \ . \eeq

\noi Actually with (\ref{2.12e}) we have written the general solution
of the inhomogeneous equation (\ref{2.2e}), the only assumption being
that the scale $Q_0^2$ allows us to define a perturbative and a
non-perturbative component. However this way of defining a
non-perturbative component is too naive and factorization scheme
dependent. Indeed let us consider the contribution of the HO inhomogeneous
kernel ${\alpha \over 2 \pi} {\alpha_s(k^2) \over 2 \pi}
k_{q}^{(1)}$ to the anomalous component (\ref{2.10e})

\beq \label{2.13e} q_{AN}^{NS}(Q^2,Q_0^2) = \cdots - \left [ 1 - \left
( {\alpha_s(Q^2) \over \alpha_s (Q_0^2}\right )^{-2
P_{qq}^{(0)}/\beta_0} \right ] {\alpha \over 2 \pi} \ {k_q^{(1)} \over
P_{qq}^{(0)}} \sigma^{NS} + \cdots \eeq

\noi where we used the running coupling constant defined by

\beq \label{2.14e} {\partial \alpha_s(Q^2) \over \partial \log Q^2} = -
\alpha_s \left ( {\alpha_s \over 4 \pi} \beta_0 + \left ( {\alpha_s
\over 4 \pi}\right )^2 \beta_1 \right ) \ . \eeq

\noi When similar expressions for $q_f^{(+)}$ are introduced in
(\ref{2.7e}), one obtains for ${\cal F}_2^{\gamma}$ a contribution
proportional to $\alpha_s^0$ (besides the Leading Logarithm contribution)

\beq \label{2.15e} {\cal F}_2^{\gamma} \sim - {\alpha \over 2 \pi} \
{k_q^{(1)} \over P_{qq}^{(0)}} \ {<e^4> \over <e^2>} + C_{\gamma} \eeq

\noi which is factorization scheme independent, and a contribution
which verifies the homogeneous LO DGLAP equation

\beq \label{2.16e} {\cal F}_2^{\gamma} \sim \sum_f e_f^2 \left [
{\alpha \over 2 \pi} \ {k_q^{(1)} \over P_{qq}^{(0)}}\ {e_f^2 \over
N_f <e>^2} +
q_{H,f}^{(+)}(Q_0^2) \right ] \left ( {\alpha_s (Q^2) \over \alpha_s (Q_0^2) }
\right )^{-2P_{qq}^{(0)}/\beta_0} \ , \eeq

\noi which must be also scheme independent. Now it is clear from
(\ref{2.16e}) that $q_{H,f}^{(+)}$ is not scheme invariant with respect to the
``photon factorization'' scheme which defines the inhomogeneous kernel
$k_q^{(1)}$. Therefore it cannot be, for instance, the same in the
$\overline{MS}$
scheme or the $DIS_{\gamma}$ scheme defined in \cite{13r}. Of course,
$q_{H}^{(+)}$ is also non-invariant with respect to the usual hadronic
factorization scheme which defines $P_{ij}^{(1)}$. Thus the assumption that the
hadronic input could be described by a VDM-type input is
clearly too naive. It may however be true in a specific factorization
scheme and we explore this possibility in the next section and in appendix A.

\mysection{The Non-Perturbative Input at Lowest Order}
\hspace*{\parindent} In order to better understand the content of
$q_{H}(Q_0^2)$, let us consider the lowest order contribution to
${\cal F}_2^{\gamma}$ coming from the imaginary part of the box
diagram, Fig.~1, which shows how the virtual photon $q$ probes the
quark content of the real photon $p$. The lower part $G(k,p)/(-k^2)$ (it
includes the quark propagators) represents the coupling of
the real photon to a $q\bar{q}$ pair and includes non-perturbative
effects.\par

\begin{figure}[htb]
\vspace{9pt}
\centering
\includegraphics[width=5in,height=2in]{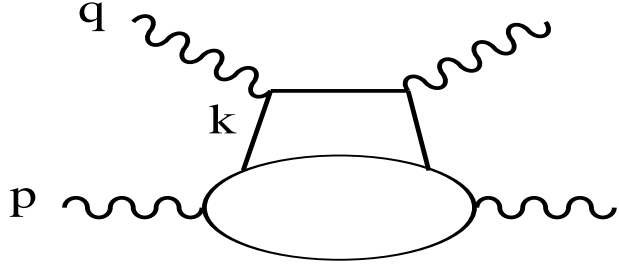}
\caption{The box diagram}
\label{fig:1}
\end{figure}

Actually our only assumption is that $G(k,p)$ tends to the pointlike
term for large $|k^2|$.

\beq \label{3.1e} \lim_{|k^2|\gg \Lambda^2} G(k,p) = G^P(k,p) \sim
\delta \left ( (p - k)^2\right ) \left [z^2 + (1 - z)^2\right ]  \eeq

\noi where $z$ is the fraction of the longitudinal $p$ momentum carried
away by $k$. When $k^2$ goes to zero, $G(k,\rho )/(-k^2)$ must be
integrable, because ${\cal F}_2^{\gamma}$ is a physical finite quantity. This
means that we must have $\lim\limits_{k^2 \to 0} G(k,p) \sim
(-k^2)^{\alpha}$ with $\alpha > 0$.\par

We make the pointlike content of $G(k,p)$ explicit by writing

\beq \label{3.2e} {G(k,p) \over - k^2} = {G^P(k,p) \over - k^2} +
{G(k,p) - \theta
(|k^2| - Q_0^2) G^{P}(k,p) \over - k^2}\ - \ {\theta (Q_0^2 - |k^2|)
G^P(k,p) \over - k^2} \
.\eeq

\noi The first term on the RHS of (\ref{3.2e}), without cut on $k^2$,
corresponds to the perturbative expression of the box diagram. \par

Its contribution, in the collinear approximation, is easily calculated
\cite{6r} and is equal to

$${\cal F}_2^{\gamma} \sim 3 e_f^4\ {\alpha \over
\pi} \left [ (x^2 + (1 - x)^2) \left ( - {1 \over \bar{\varepsilon}}
\left (   {Q^2 \over \mu^2}\right )^{- \varepsilon}\right ) + (x^2 + (1 -
x)^2) \ln (1 - x) + 2x(1- x) \right ]$$
\beq \label{3.6e}
\equiv 3e_f^4 \ {\alpha \over \pi} \left ( - {1 \over
\bar{\varepsilon}} + \ln {Q^2
\over \mu^2} \right ) (x^2 + (1 - x)^2) + C_{\gamma , c}^f \eeq

\noi in which we add a quark $f$ and an antiquark $\bar{f}$ contribution.
We define $Q^2 = - q^2$ and $x = Q^2/2p\cdot q$. Note that we took
the upper bound of the integral over
$|k^2|$ equal to $Q^2$. (The actual bound is $Q^2/x$, but this
$x$-dependence is beyond the collinear approximation). \par

This expression for the box diagram has been obtained with the
dimensional regularization $({1 \over \bar{\varepsilon}} = {1 \over
\varepsilon} - \gamma_E + \ln \ 4 \pi)$; it is the one used to
define the $\overline{MS}$ factorization scheme which consists in
subtracting the term proportional to
$(Q^2/\mu^2)^{-\varepsilon}/\bar{\varepsilon}$. This procedure defines
the scheme-dependent direct term $C_{\gamma , c}^f$ in the collinear
approximation (or $C_{\gamma}$ given in (\ref{2.8e}) when we take into
account the non-collinear terms).\par

${\cal F}_2^{\gamma}(x)$ being a physical quantity, it cannot contain
the $1/\varepsilon$ pole, and it is here that the third term of the RHS
of (\ref{3.2e}) plays its part. We obtain from this last term

\beq \label{3.7e}
   {\cal F}_2^{\gamma ,c}       = - 3e_f^4 \ {\alpha \over
\pi} \left ( - {1 \over \bar{\varepsilon}} + \ln {Q_0^2 \over \mu^2}
\right ) (x^2 + (1 - x)^2) - C_{\gamma , c}^f \ . \eeq

This term has no anomalous $\ln Q^2$ behavior. Actually it is
independent of $Q^2$, when QCD is not switched on. When the part of
${\cal F}_2^{\gamma ,c}$ proportional to
$(-{1 \over \bar{\varepsilon}} + \ln {Q_0^2 \over \mu^2})$ is added
to (\ref{3.6e}), the $1/\bar{\varepsilon}$ poles cancel each other and $\ln
     Q^2/\mu^2$ is changed into

\beq \label{3.8e} \ln \ {Q^2 \over Q_0^2} = \ln \ {Q^2 \over
\Lambda^2} - \ln\ {Q_0^2 \over \Lambda^2} = {4 \pi \over \beta_0
\alpha_s(Q^2)} \left ( 1 - \left ( {\alpha_s(Q^2) \over
\alpha_s(Q_0^2)}\right ) \right ) \ . \eeq

\noi This $Q^2$-dependence corresponds to the LO part of (\ref{2.10e}) with
$P_{qq}^{(0)} = 0$. \par

Let us now consider the second term of (\ref{3.2e}). The
$\theta$-function cuts the $1/k^2$ perturbative behavior of this
contribution. The integration over $k^2$ is therefore controlled by the
non-perturbative behavior of $G(k,p)/(-k^2)$ and we obtain a result which
does not depend on $Q^2$. The value of $Q_0^2$ must of course be
chosen such as
$G^{NP}(p,k,Q_0^2) \equiv G(k,p) - \theta (|k^2|- Q_0^2)G^P(k,p)$
represents the
non-perturbative input. For instance an overly large value of $Q_0^2$ would
conduce to a perturbative tail in $G^{NP}(p,k,Q_0^2)$. \par

We now define the non-perturbative quark content of the real photon by

\beq
\label{3.9e}
\int d^4k \ \delta \left ( z - {k\cdot n \over p\cdot n}\right )
{G^{NP}(p,k,Q_0^2) \over (-k^2)} = q^{NP}(z, Q_0^2)
\eeq

\noi ($n$ is a light-cone vector such as $k \cdot n \sim k^0 +
k^z$). With (\ref{3.9e}) we have defined a non-perturbative input (if
$Q_0^2$ is correctly chosen) which is {\it invariant with respect to
the photon factorization scheme}.  Indeed it does not depend on the
regularization used to calculate (\ref{3.6e}) nor on the subtraction
defining the $\overline{MS}$ scheme. When the QCD evolution is switched
on (all order QCD expressions are discussed in appendix A), both
$q^{NP}$ and $C_{\gamma , c}^f$ acquire a hadronic
$Q^2$-dependence and we obtain a hadronic contribution (which behaves
like a hadronic structure function) to ${\cal F}_2^{\gamma}$

\beq \label{3.10e} q_f^{H}(Q^2) = \left ( {\alpha_s(Q^2) \over
\alpha_s(Q_0^2)}\right )^{-2P_{qq}^{(0)}/\beta_0} \left ( q_f^{NP}
(Q_0^2) - {C_{\gamma ,
c}^f \over 2e_f^2}\right ) \ . \eeq

         This hadronic contribution is scheme dependent because of the
presence of $C_{\gamma,c}^f$, but $q_f^{NP}(Q_0^2)$ is not. Therefore,
in the $\overline{MS}$ factorization scheme, the hadronic input is
given by expression (\ref{3.10e}), and, at $Q^2 = Q_0^2$, we have

\beq \label{3.11e} {\cal F}_2^{\gamma}(x, Q_0^2) = C_{\gamma}(x) +
\sum_{f=1}^{N_f} \left [ e_f^2\left ( q_f^{NP} (Q_0^2) +
\bar{q}_f^{NP}(Q_0^2)\right ) - C_{\gamma , c}^f \right ] \ . \eeq

In the above expression, we only studied the part of ${\cal
F}_2^{\gamma}$ associated with the quark contributions. (Nor did we
write the convolution with the Wilson coefficient). Similar
considerations applied to the gluon distribution lead to modifications
of the input starting at order ${\cal O}(\alpha_s)$. The NNLO
corrections are not considered in this paper.\par

These results are different from those obtained by the authors of ref.
\cite{13r,7r,8r} who work in a factorization scheme called
$DIS_{\gamma}$ in which

\beq \label{3.12e} {\alpha \over 2 \pi} \ {<e^4> \over <e^2>}\
k_q^{(1)} (DIS_{\gamma}) = {\alpha \over 2 \pi} \ {<e^4> \over <e^2>}\
k_q^{(1)}(\overline{MS}) - C_{\gamma} P_{qq}^{(0)} \ , \eeq

\noi so that $C_{\gamma}(DIS_{\gamma}) = 0$. In this case, the structure
function is written

\beq
\label{3.13e}
{\cal F}_2^{\gamma}(x,Q_0^2) = \sum_{f=1}^{N_f} e_f^2 \left (
q^{NP}_{f, DIS_{\gamma}} (Q_0^2)+ \overline{q}^{NP}_{f, DIS_{\gamma}}
(Q_0^2)\right )
\eeq

\noi where $q^{NP}_{f, DIS_{\gamma}} (Q^2)$ is the non-perturbative
input in this particular factorization scheme. \par

Let us finish this discussion by emphasizing the fact that the parton
distributions defined by (\ref{2.12e}) and (\ref{3.10e}) are universal
(independent of the particular reaction studied here, namely the DIS on
a real photon). Of course they are factorization scheme dependent and
here we work in the $\overline{MS}$ scheme.

\mysection{The Vector Dominance Model}
\hspace*{\parindent}
The non-perturbative contribution defined in (\ref{3.9e}) is not known.
This is why we could proceed as in the pure hadronic case by defining a
parameter-dependent input and by determining the parameters by a fit to
data. Here we prefer to follow another path and to try to constrain the
non-perturbative input by assuming that it can be described by the
quark and gluon distributions in Vector Mesons. This assumption, the
Vector Dominance Model, is known to work well in the non-perturbative
domain and to correctly describe how photons couple to quarks. We used
this assumption in our preceding paper \cite{6r}, which led to the
AFG parametrization. Here we keep this approach, but we
make it more flexible by
varying the non-perturbative normalization of the gluon and sea quarks.
We also consider modifications of the gluon $x$-shape. \par

In ref. \cite{6r} we considered the photon as a coherent superposition of
vector mesons

\beq \label{4.1e} \gamma = {g \over \sqrt{2}} \left ( \rho + {\omega
\over 3} - {\sqrt{2} \over 3} \phi \right ) = g \left ( {2 \over 3}
u\bar{u} - {1 \over 3} d\bar{d} - {1 \over 3} s\bar{s}\right ) \eeq

\noi with a coupling constant $g$ determined from the
$\sigma_{tot}(\gamma p)$ and $\sigma_{tot}(\pi p)$ cross sections

\beq \label{4.2e} g^2 \simeq \alpha \ . \eeq

\noi Assuming that the parton distributions in the $q\bar{q}$ ``bound
states'' of (\ref{4.1e}) are similar to those of the pion, observed in
Drell-Yan and direct photon reactions \cite{11r}, we can write
\bminiG{4.3e} \label{4.3ae} u_{valence}^{\gamma} (x, Q^2) = g^2 \ {4
\over 9} \ u_{valence}^{\pi} \eeeq
\beeq \label{4.3be}
u_{sea}^{\gamma}(x, Q^2) = g^2 \left ( {4 \over 9} + {1 \over 9} + {1
\over 9}\right )
u_{sea}^{\pi} = g^2 {2 \over 3} u_{sea}^{\pi} \eeeq
\beeq \label{4.3ce} G^{\gamma}(x, Q^2) = g^2 {2 \over 3} g^{\pi} (x,
Q^2)
\emini

\noi and so on for the parton distributions of the non-perturbative
component of the real photon (we assume a SU(3) flavor symmetry). \par

This rough approach leads to a reasonable agreement with data \cite{6r}.
Here we would like to make it more flexible. Let us start from the
parametrization of the pion structure function at $Q^2 = 2$~GeV$^2$
taken from ref.
\cite{11r} and used in ref. \cite{6r}
\bminiG{4.4e} \label{4.4ae}
xu_{valence}^{\pi}  = C_v\ x^{p_2} (1 - x)^{p_3} \quad ( p_2 =
.48\ , \ p_3 = .85) \eeeq \beeq \label{4.4be} xq_{sea}^{\pi} = C_s(1 -
x)^{p_8} \quad (p_8 = 7.5\ , \ C_s = 1.2) \eeeq \beeq \label{4.4ce}
xg^{\pi} = C_g(1 - x)^{p_{10}} \quad (p_{10} = 1.9) \emini

\noi with $C_v = 1/B (p_2, 1 + p_3)$ ($B(x,y)$ is the
beta-function) and $C_g
= (1 + p_{10})(1 - {C_s \over 1 + p_8} - {2 p_2 \over 1 +
p_3 + p_2}) = .447(1 + p_{10})$, $C_g$ being determined in
such a way that

\beq
\label{4.5e}
\int_0^1 dx \ x \left [ 2 u_{valence}^{\pi} + q_{sea}^{\pi} + g^{\pi}
\right ] = 1 \ .
\eeq

\noi This input is in good agreement with ${\cal
F}_2^{\gamma}(x, Q^2)$ data which mainly constrain the quark distributions.
Therefore we leave the quark distributions fixed and we take
$p_{10}$ as a free parameter. As we shall see below, ${\cal
F}_2^{\gamma}$ is not sensitive to variations of $p_{10}$, a
parameter which should be constrained by photoproduction data. We also
leave some freedom in the normalization of the distributions. First of
all the overall normalization is allowed to vary around the value fixed in
(\ref{4.2e}) and we write

\beq
\label{4.6e}
g^2 = C_{np} \cdot \alpha \ .
\eeq

Then we also consider the possibility of having a different coupling of
the photon to the valence distributions and to the sea and gluon
distributions (this extra coupling could proceed through a quark loop).
We parametrize this possibility by a modification of $C_s$ and $C_g$

\bminiG{4.7e} \label{4.7ae} C_s = C_{mom} \cdot  1.2  \eeeq \beeq \label{4.7be}
         C_g = C_{mom} \cdot 0.447 \ (1 + p_{10}) \ ,\emini

\noi the default value being $C_{mom} = 1.0$. \par

Let us end this section by discussing another input, the quark
masses. Threshold effects due to the charm quark may be important in
${\cal F}_2^{\gamma}(x, Q^2)$ at large $x$. To study this problem, let
us again consider the box diagram and the massive quark contribution to
${\cal F}_2^{\gamma}$. Dropping all inessential factors and neglecting
terms of order ${\cal O}(m^2/Q^2)$, one obtains \cite{new23}

$${\cal F}_{2,m}^{\gamma} \sim \theta (1 - \beta ) \Big \{ (x^2 + (1
- x)^2) \ln {Q^2 \over m_q^2} + (x^2 + (1-x)^2) \ln \left (
\left ( {1 + \sqrt{1 - \beta} \over 2}\right )^2 {s \over Q^2}\right
)$$
\beq
\label{4.8e}
   + (8x (1 - x) - 1) \sqrt{1 - \beta} \Big \} \ ,
\eeq

\noi with $s = (p+q)^2$ and $\beta = 4m_q^2/s$. Subtracting the term
proportional to $\log {Q^2 \over m_q^2}$, we define a massive direct
term

$$C_{\gamma}(x, m_q) = e_c^4 {\alpha \over \pi} 3 \theta (1 - \beta )
\Big [ (x^2 + (1 - x)^2) \ln \left ( \left ( {1 - x \over x}
\right ) \left ( {1 + \sqrt{1 - \beta} \over 2}\right)^2 \right )$$
\beq
\label{4.9e}
+ (8 x (1 - x) - 1) \sqrt{1 - \beta }\Big ]
\eeq

\noi which has the massless limit (\ref{2.8e}) when $\beta = 0$.\par

Then the term proportional to $\ln {Q^2 \over m_q^2}$ is replaced by
the one generated by the massless evolution equations (\ref{2.1e}) and
(\ref{2.2e}) with the boundary conditon $q(x,Q^2 = m_q^2) = 0$. This
evolution exactly reproduces, at the lowest order in $\alpha_s$, the
$\ln {Q^2 \over m_q^2}$-term of (\ref{4.8e}). Therefore, close to
the threshold $Q^2 = m_q^2$, ${\cal F}_2^{\gamma}$ of expression
(\ref{2.7e}) reproduces the behavior (\ref{4.8e}) of the box diagram
contribution. Let us also note that we keep in $C_{\gamma}(x,
m_q)$ the terms of order ${\cal O}(m^2/Q^2)$ when $Q^2 \sim m_q^2$.\par

The effect of the massive direct term (\ref{4.9e}) is important when
$x$ goes to $x_{th} = 1/(1 + 4m_q^2/Q^2)$. One then gets $\ln {(1 +
\sqrt{1 - \beta})^2 \over \beta} \sim 2 \sqrt{1 - \beta}$ and
${\cal F}_{2,q}^{\gamma}$ given by (\ref{4.8e}) goes to zero, whereas the use
of the massless limit (\ref{2.8e}) (without cut on $x$) leads to a
negative contribution when $x$ goes to 1. \par

However one must keep in mind that in most applications, we are far
from the threshold and the massless evolution of the charm distribution
is a good approximation which allows us to take into account the effects
of the QCD evolution, not present in (\ref{4.8e}). But this is not true
for the bottom distribution as long as $m_b^2/Q^2$ is large. The
distributions presented in this paper are obtained by solving eq. (\ref{2.1e})
and (\ref{2.2e}) with $N_f = 3$ for $Q_0^2 \leq Q^2 \leq m_c^2$, $N_f = 4$ for
$m_c^2 < Q^2 < m_b^2$ ($m_c = 1.41$~GeV) and $N_f = 5$ for $m_b^2 <
Q^2$ ($m_b = 4.5$~GeV).

\section{Analysis of LEP data}
\hspace*{\parindent}
In this section we analyze data on ${\cal F}_2^{\gamma}$ in the light of the
parametrization discussed in sections 3 and 4. Once we assume that the
non-perturbative input can be determined within the framework of the Vector
Dominance Model as explained in the preceding section, the number of
free parameters is considerably reduced. ${\cal F}_2$ is barely sensitive to
the gluon distribution parameters $C_{mom}$ and $p_{10}$ that we
shall discuss in relation to photoproduction reactions and, for the
time being, we keep these parameters equal to their default values
$C_{mom} = 1.0$ and $p_0 = 1.9$. Therefore only two free parameters
remain, $C_{np}$ which fixes the overall normalization of the non
perturbative input (expression (\ref{4.6e})) and $Q_0^2$ which fixes the
boundary between the perturbative and the non-perturbative model.\par

If data were precise enough, it should be possible to constrain
$C_{np}$ and $Q_0^2$ separately. In fact the VDM contribution decreases
rapidly with $x$ and the perturbative contribution is dominant at large
$x$. Therefore in this kinematical domain, it should be possible to
determine $Q_0^2$ by the means of medium $Q^2$ data (indeed for overly large
$Q^2$, ${\cal F}_2^{\gamma}$ is no longer sensitive to $Q_0^2)$. At small
values of $x$ on the contrary the non-perturbative input is large and
data should constrain $C_{np}$. As we shall see this ideal situation is
not realized, because the data existing at large values of $x$ are very
poor, and we are led to fit low-$x$ data where $C_{np}$ and
$Q_0^2$ are correlated.\par

First let us concentrate on LEP data at small and medium $Q^2$, an
overall comparison with all existing data will be conducted at the end of this
section. The data that we analyze are in the range $3.7 \leq Q^2 \leq
17.3$~GeV$^2$ and belong to the four LEP experiments. ALEPH \cite{16r}
LEP2 data at 17.3~GeV$^2$, DELPHI \cite{17r} LEP1 data at 12.7 GeV$^2$
and 5.2 GeV$^2$, L3 \cite{18r} LEP2 data at 15.3 GeV$^2$, and OPAL
\cite{19r} LEP1 ($Q^2 = 3.7$~GeV$^2$) and LEP2 ($Q^2 = 10.7$~GeV$^2$)
data. The comparison between NLO theory and data is done in Figs. 2 and
3 where we show the theoretical curves obtained for $Q_0^2 =
.3$~GeV$^2$ and for $Q_0^2 = 1.0$~GeV$^2$ with $C_{np} = 1$. We work in
the $\overline{\rm MS}$ scheme and use $\Lambda_{\overline{MS}}^{(4)} =
300$~MeV, a value which is in agreement with the world average and a
determination obtained by fitting photon structure function data
\cite{Z}. The coupling $\alpha_s(Q^2)$ is obtained by exactly solving
eq. (\ref{2.14e}).\par

\begin{figure}[htb]
\vspace{9pt}
\centering
\includegraphics[width=3in,height=2.5in]{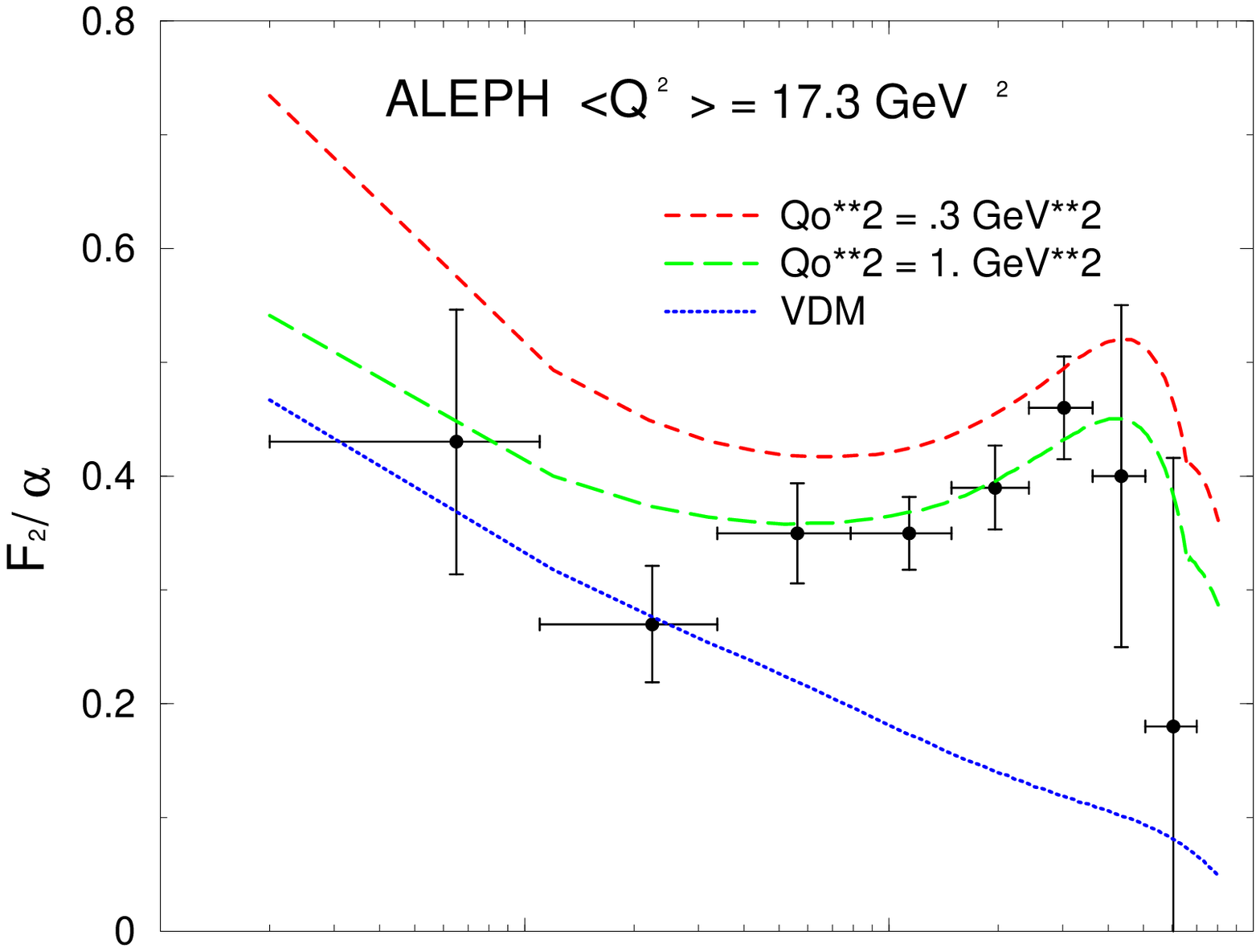}
\includegraphics[width=2.8in,height=2.5in]{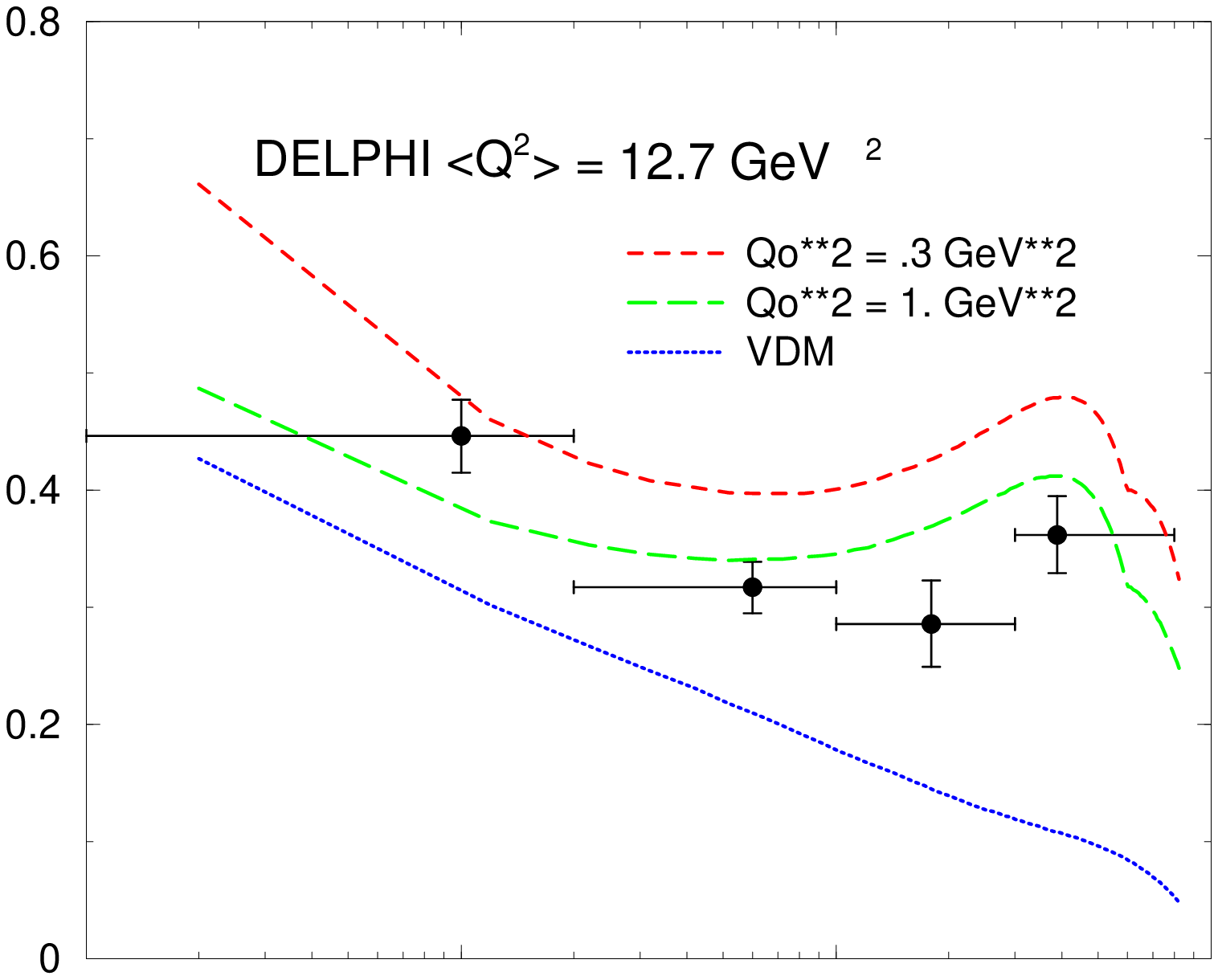}
\includegraphics[width=3in,height=2.5in]{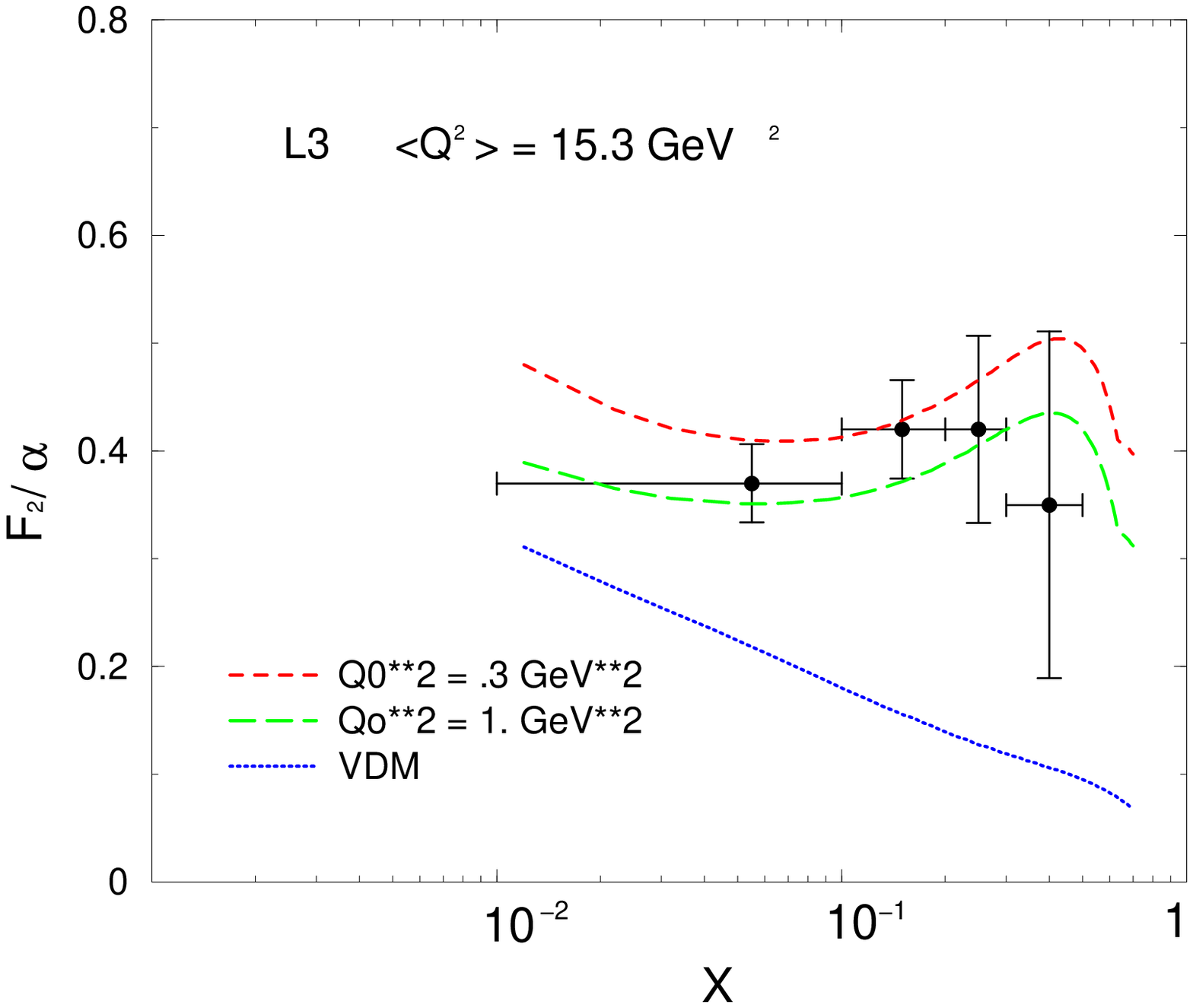}
\includegraphics[width=2.8in,height=2.5in]{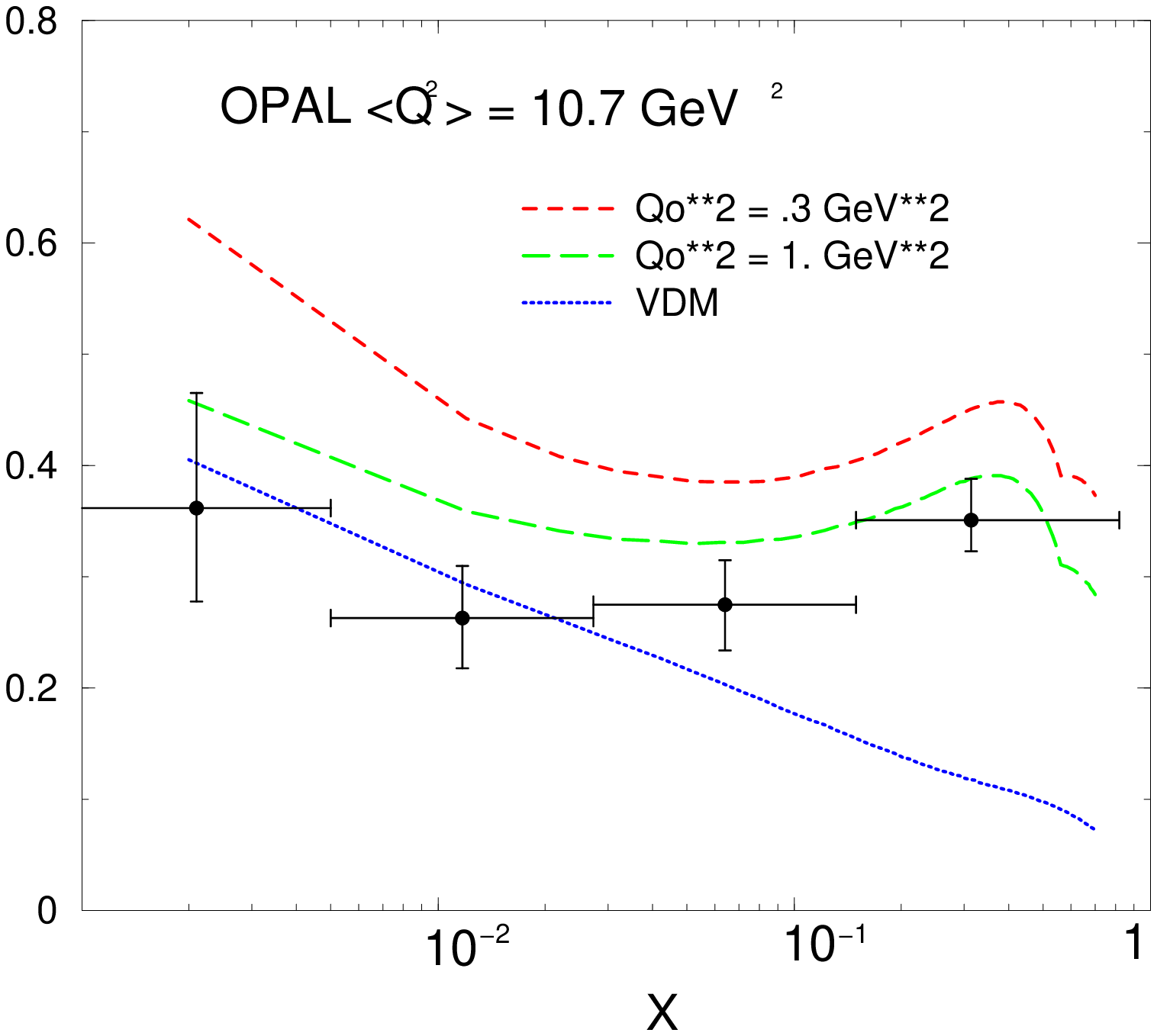}
\caption{Theory confronted with ALEPH \cite{16r}, DELPHI \cite{17r},
L3 \cite{18r} and OPAL \cite{19r} data in the range 10 GeV$^2 \ \lsim
\ <\! Q^2\!> \ \lsim$ 18~GeV$^2$.}
\label{fig:2}
\end{figure}

\begin{figure}[htb]
\vspace{9pt}
\centering
\includegraphics[width=2.8in,height=2.8in]{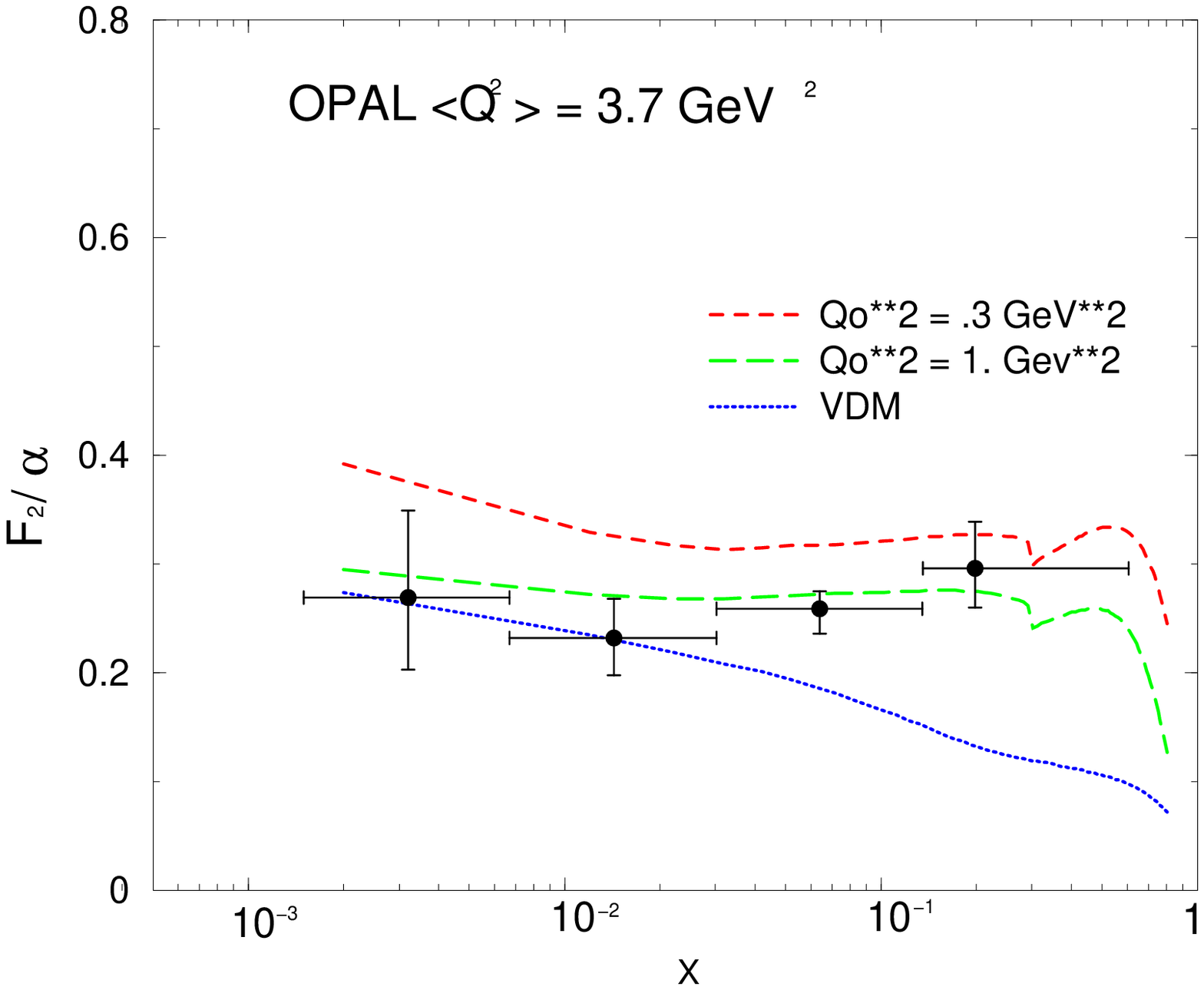}
\includegraphics[width=2.8in,height=2.8in]{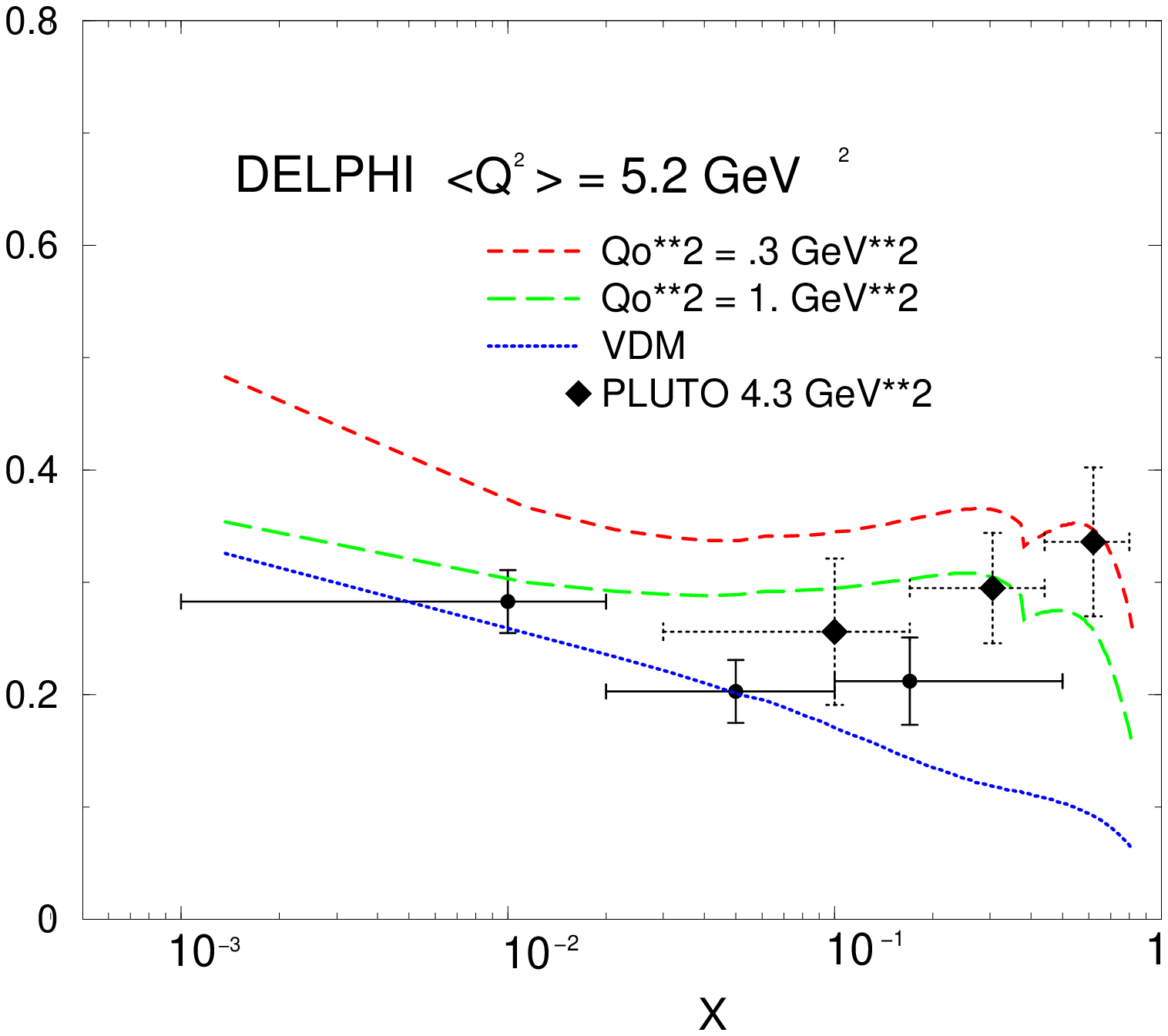}
\caption{Theory confronted with low-$Q^2$ data from OPAL, DELPHI and
PLUTO \cite{20r}.}
\label{fig:3}
\end{figure}

In Fig. 2, we see that large-$x$ data are poor. Either they have large
error bars (ALEPH, L3), or they correspond to large $x$-bins (DELPHI,
OPAL)\footnote{The errors are the total errors (when given by the
experiments), or the linear sum of the syst. and stat. errors (OPAL).
We also investigated the effect of quadratically summed errors in the
fit performed below. Our best fit parameters change by less than
10~\%. Correlations between data points are not taken into account.}.
Therefore they cannot accurately constrain $Q_0^2$ and do not allow us
to check the predicted $x$-dependence of $F_2^{\gamma}(x, Q^2)$. On the
other hand, all large-$x$ data points ($x \ \gsim\ .2$) favor a value
of $Q_0^2$ close to 1 GeV$^2$~; the choice $Q_0^2 = .3$~GeV$^2$ leads
to predictions well above the experimental points. This is also true
for smaller values of $x$ where the VDM contribution is large. However
in this $x$-region we expect a strong correlation between $Q_0^2$ and
$C_{np}$.\par

The low-$Q^2$ data of Fig.~3 lead us to the same conclusion. However
whereas the agreement between OPAL data and theory is good, suggesting
a slightly smaller contribution of the VDM component, DELPHI data on
the contrary suggest a very large suppression, by some 30~\%, of the VDM
component. It is worth noting that the DELPHI points at $x = .05$ and
$x = .17$ are below the corresponding OPAL points, although the $Q^2$
value is larger. Other data at low $Q^2$ do not clarify the situation.
PLUTO data \cite{20r} are close to the OPAL points, but the error bars
are very large. L3 data \cite{21r} ($Q^2 = 5$~GeV$^2$, not shown) are
noticeably above OPAL data at small-$x$, $x \sim 10^{-2}$, and
TPC-2$\gamma$ \cite{22r} data at $Q^2 = 5.1$~GeV$^2$ (not shown) have
large error bars.\par

In order to determine the values of $Q_0^2$ and $C_{np}$, we proceed to
a best fit of the LEP data displayed in Figs.~2 and 3. The theoretical
values are calculated with all parameters kept fixed, except $Q_0^2$
and $C_{np}$, and they are averaged over the corresponding
experimental
$x$-bins. For a given value of $Q_0^2$, we look at the value of $C_{np}$
which minimizes the $\chi^2$ value, and we obtain the results shown in
Fig.~4 for the ALEPH and DELPHI experiments. At the minimum of the curves, the
corresponding values of the non-perturbative normalization are
respectively $C_{np}$=.60 and $C_{np}$=1.05.

\begin{figure}[htb]
\vspace{9pt}
\centering
\includegraphics[width=3in,height=2.6in]{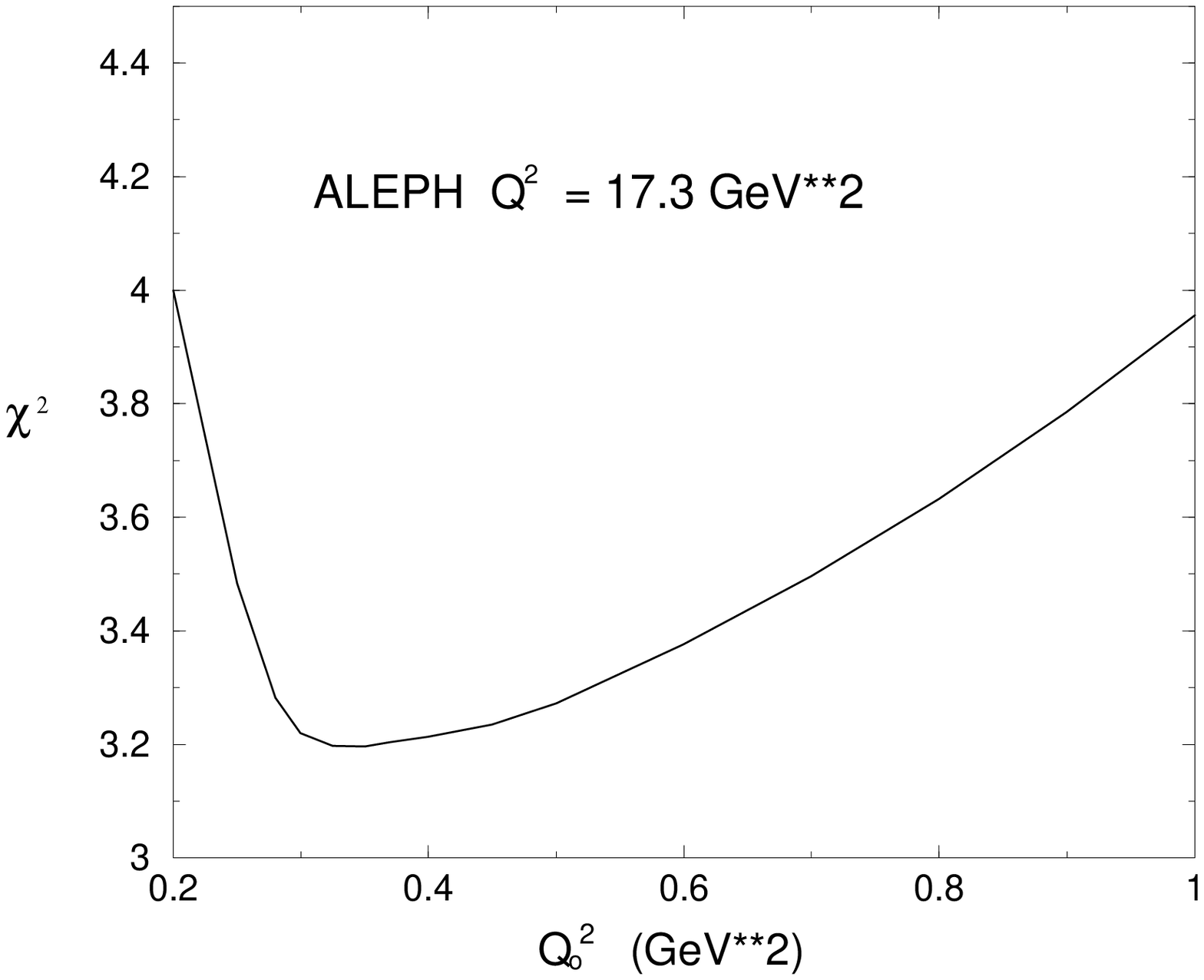}
\includegraphics[width=2.8in,height=2.6in]{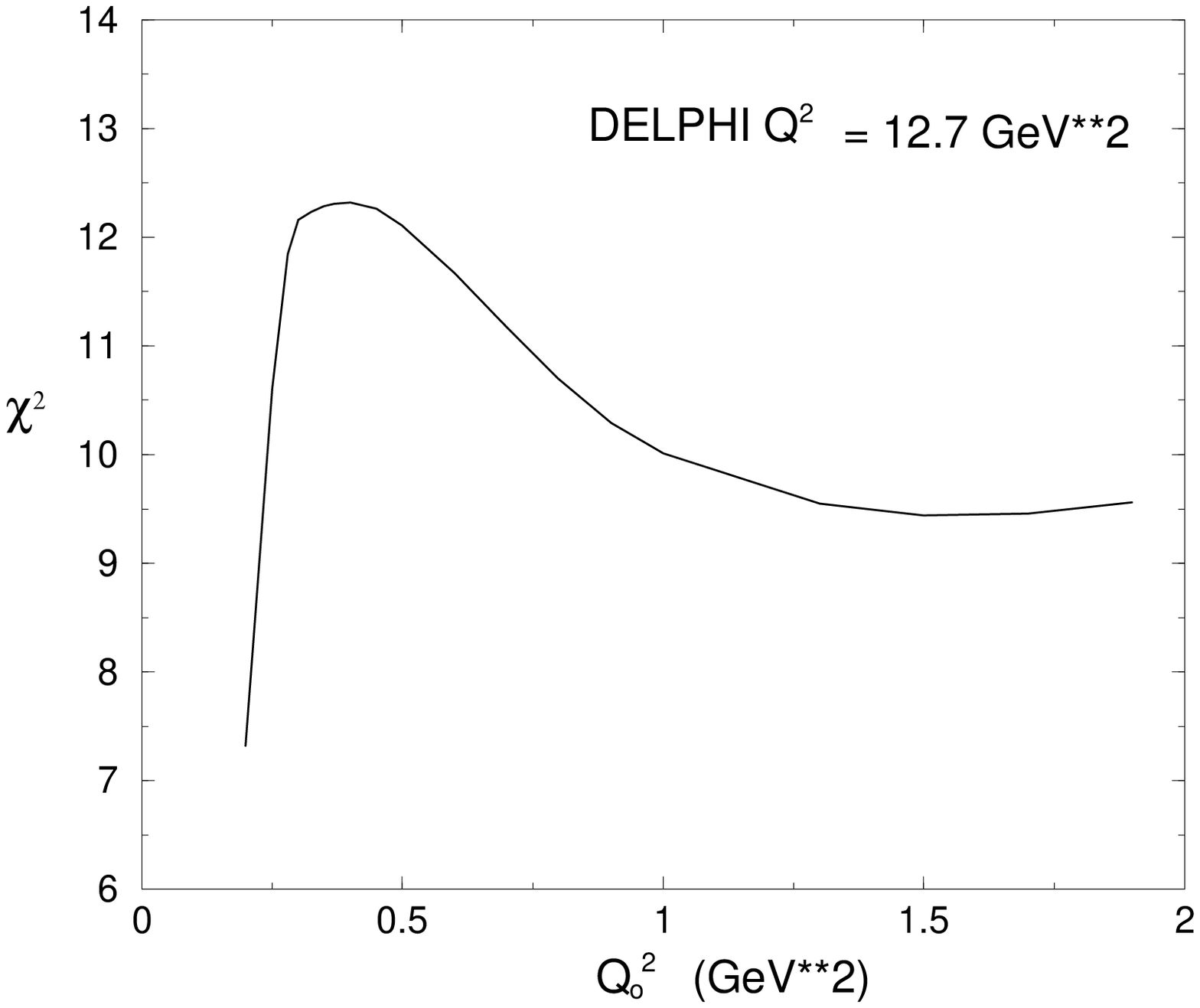}
\caption{$\chi^2$-values as a function of $Q_0^2$}
\label{fig:4}
\end{figure}

We obtain very different shapes of the $\chi^2$-curves. In
fact, L3 and OPAL data lead to $\chi^2$-curves very similar to the one
displayed in Fig.~4 for ALEPH. Only DELPHI data lead to a false minimum
at  $Q_0^2$=1.5 GeV$^2$ and $C_{np}$=1.05; there is no true minimum for
$Q_0^2 \geq .2$~GeV$^2$. It is easy to find the reason for this behavior. The
first point at small-$x$ of DELPHI data is high with respect to the next
point. This configuration, associated with small errors, drives the fit
to small $Q_0^2$ and $C_{np}$ values and reproduces the steep
slope of $F_2^{\gamma}(x,Q^2)$ at small-$x$. Such an effect is not
present in the other
sets of data. The small errors of the DELPHI data give an important
weight to this experiment in a fit of all the data sets shown in Fig.~2 and
3. Therefore when we perform such a fit, we find (Fig.~5 (left)) a result
similar to the DELPHI fit of Fig.~4.

\begin{figure}[htb]
\vspace{9pt}
\centering
\includegraphics[width=3in,height=2.6in]{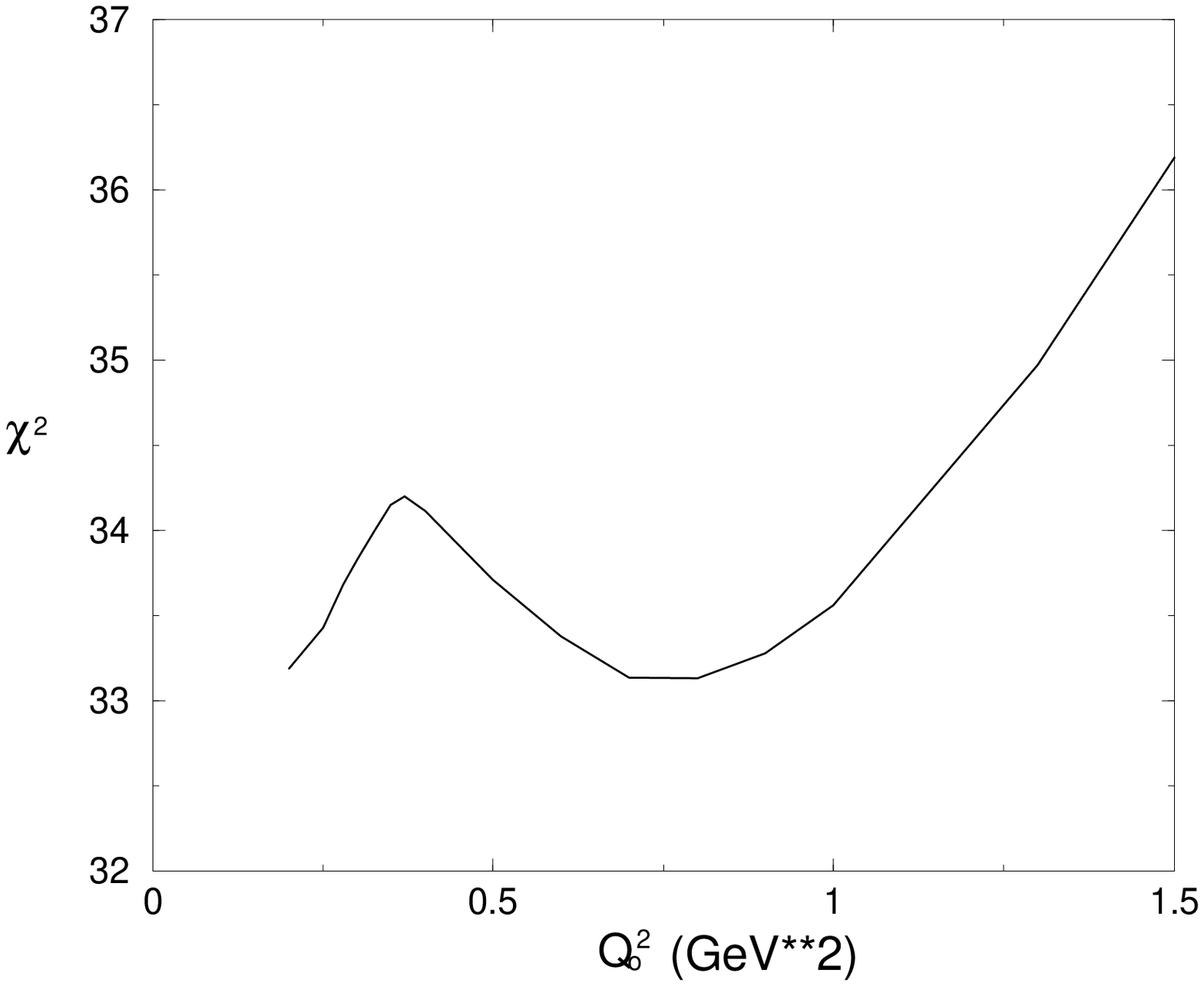}
\includegraphics[width=2.8in,height=2.6in]{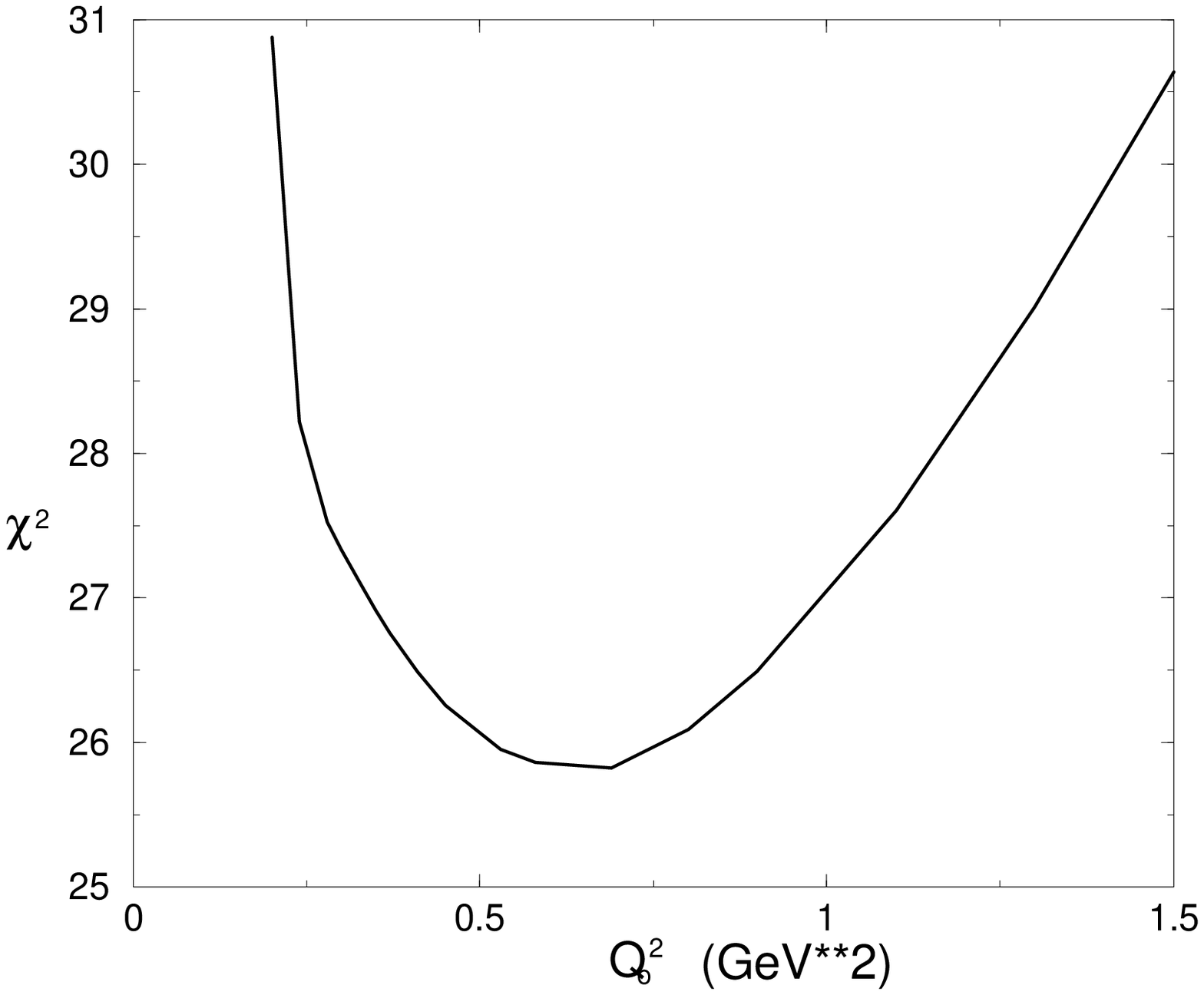}
\caption{$\chi^2$-values of the overall fit. Without modification of the
DELPHI errors (left) and with modification (right)}
\label{fig:5}
\end{figure}

As explained in the beginning of this section, we selected medium and
low-$Q^2$ LEP data in order to constrain the value of $Q_0^2$. This led
us to consider the LEP1 data from DELPHI. However in the LEP2 data from
the same experiment, the small-$x$ behavior observed in the LEP1 data is less
marked. The DELPHI collaboration, using different Monte Carlo
generators, noticed a strong model dependence of
the resulting data on $F_2^{\gamma}(x,Q^2)$. For instance the
small-$x$ data at $Q^2=19$~GeV$^2$ show a sizeable dependence on
these generators and moreover, two
(over three) of the values obtained for $F_2^{\gamma}(x, Q^2)$ are
{\bf smaller} than the one
obtained at LEP1 ($Q^2=12.7$~Gev$^2$). Therefore, in order to take
this scattering into
account, we change the errors in the
first $x$-bin of the LEP1 DELPHI preliminary data by a factor two. After this
modification, we obtain the $\chi^2$-curve shown in Fig.~5 (right). The
minimum value of the curve at $Q_0^2\simeq .7$~GeV$^2$ is obtained for
$C_{np}\simeq.78$, and the $\chi^2$ by degree of freedom is $\chi^2_{df} =
1.03$.\par

This result demonstrates the good agreement between our
theoretical input (\ref{3.11e}) and LEP data. However, as already
discussed, these data do not well constrain the large-$x$ behaviour of
$F_2^{\gamma}$ and the good $\chi^2$ shown in Fig.~5 cannot be seen as
a confirmation of our large-$x$ theoretical expression. On the other
hand we must note that some LEP data are marginally
compatible\footnote{Note also that some data are not corrected for
the limit $P^2 \to 0$  of the target photon virtuality.}. This is
shown in Fig.~6 in which we
display the $\Delta\chi^2 = 1$ contour in the $Q_0^2$ and $C_{np}$
plane. The best fits to the individual data sets are also exhibited and are
scattered outside the contour. For instance the L3 data compared to
theory calculated with the overall best fit parameters ($Q_0^2 = .7$,
$C_{np} = .78$) lead to $\chi^2 = 4.4$ which must be compared to the
value obtained at the L3 best fit point ($.7, 1.1$) $\chi^2 = .85$. For
OPAL (10.7) we obtain $\chi^2 = 1.76$ compared to $\chi^2 = .072$.\par

\begin{figure}[htb]
\vspace{9pt}
\centering
\includegraphics[width=3in,height=2.9in]{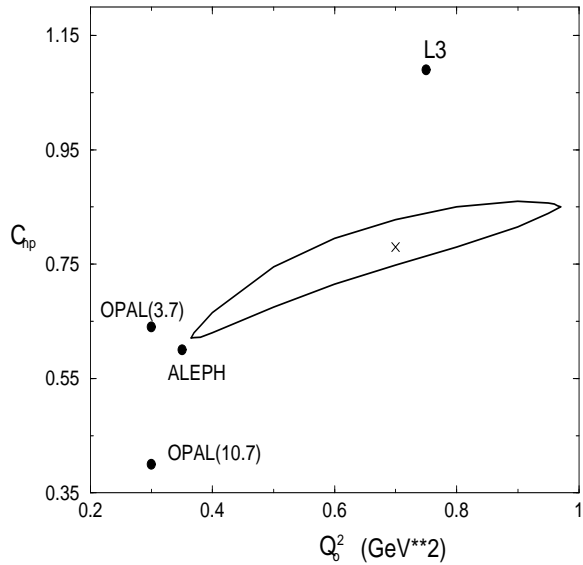}
\caption{The $\Delta \chi^2 = 1$ contour in the $Q_0^2 - C_{np}$ plane with
the individual best fits of the LEP data. The DELPHI (12.7) point
($Q_0^2 = 1.5$~GeV$^2$, $C_{np} = 1.05$) is outside the figure and DELPHI (3.7)
has no minimum for $Q_0^2 < 1.9$~GeV$^2$.} \label{fig:6}
\end{figure}

   In order to partially take into account
this scattering, we provide three parametrizations compatible with the
contour of Fig.~6, corresponding to three different values of $Q_0^2$~:
$Q_0^2 = .34$~GeV$^2$, $Q_0^2 = .70$~GeV$^2$ (best-fit parametrization) and
$Q_0^2 = .97$~GeV$^2$. We use the notation\footnote{The
parametrization AFG04(.5, 1., 1.9) has
been used in ref.
\cite{new49} under the name AFG02 and in ref. \cite{9r} under the name
AFG04.} AFG04 $(Q_0^2, C_{np}, p_{10})$ and
the short-hand notations AFG04\_BF = AFG04(.7, .78, 1.9), AFG04\_LW =
AFG04(.34, .6, 1.9) and AFG04\_HG = AFG04(.97, .84, 1.9) for the
parametrizations presented
in Appendix B. \par

We would obtain similar results from the world data on $F_2^{\gamma}$,
but with a larger scattering of the various experiments. This can be
observed from Figs.~7, 8 and 9 in which we compare our best fit
prediction with these data \cite{16r}-\cite{35NEW}. Whereas the overall
agreement is good, some data are clearly outside the general trend
represented by the best fit. Therefore a general fit leading to a
single parametrization has not a clear meaning. It is why we prefer to
``frame'' the data by several parametrizations, as we did after the
analysis of recent low and medium $Q^2$ data sensitive to $Q_0^2$ and
$C_{np}$.

\begin{figure}[htb]
\vspace{9pt}
\centering
\includegraphics[width=5.7in,height=5.2in]{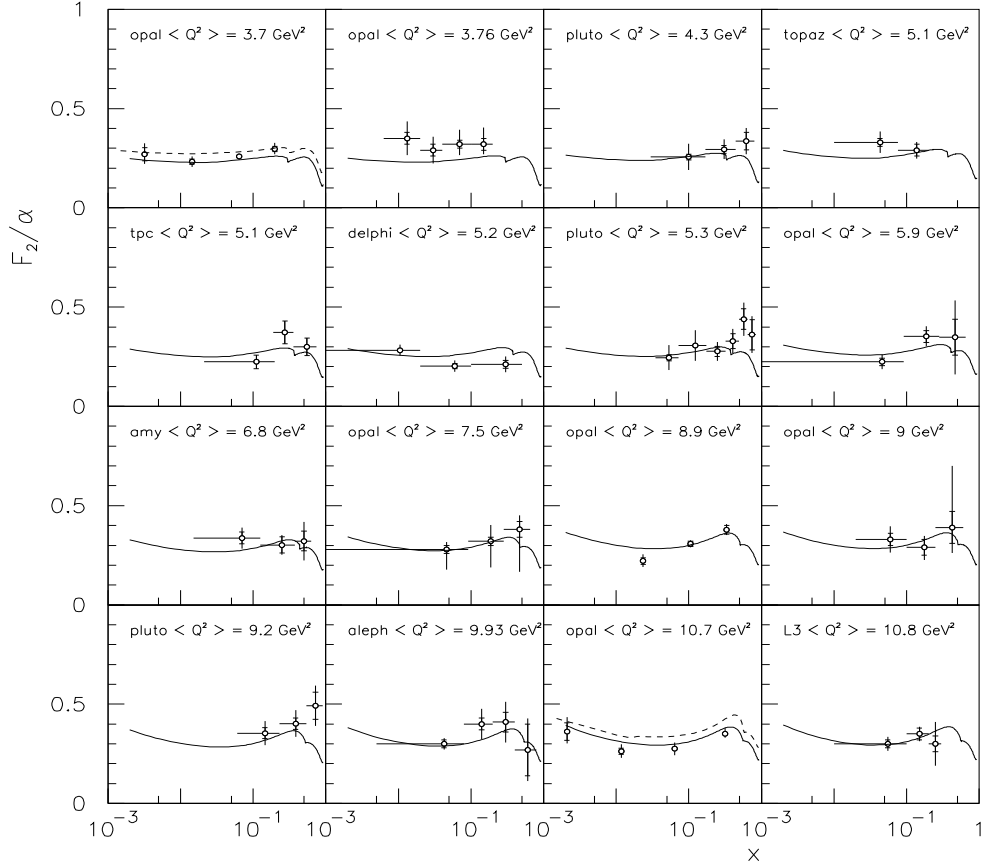}
\caption{Comparison between the best fit structure function and the
world data with 3.7 GeV$^2 < \ \ <\!Q^2\!>\  \ < 11$~GeV$^2$. }
\label{fig:7}
\end{figure}

\begin{figure}[htb]
\vspace{9pt}
\centering
\includegraphics[width=5.7in,height=5.2in]{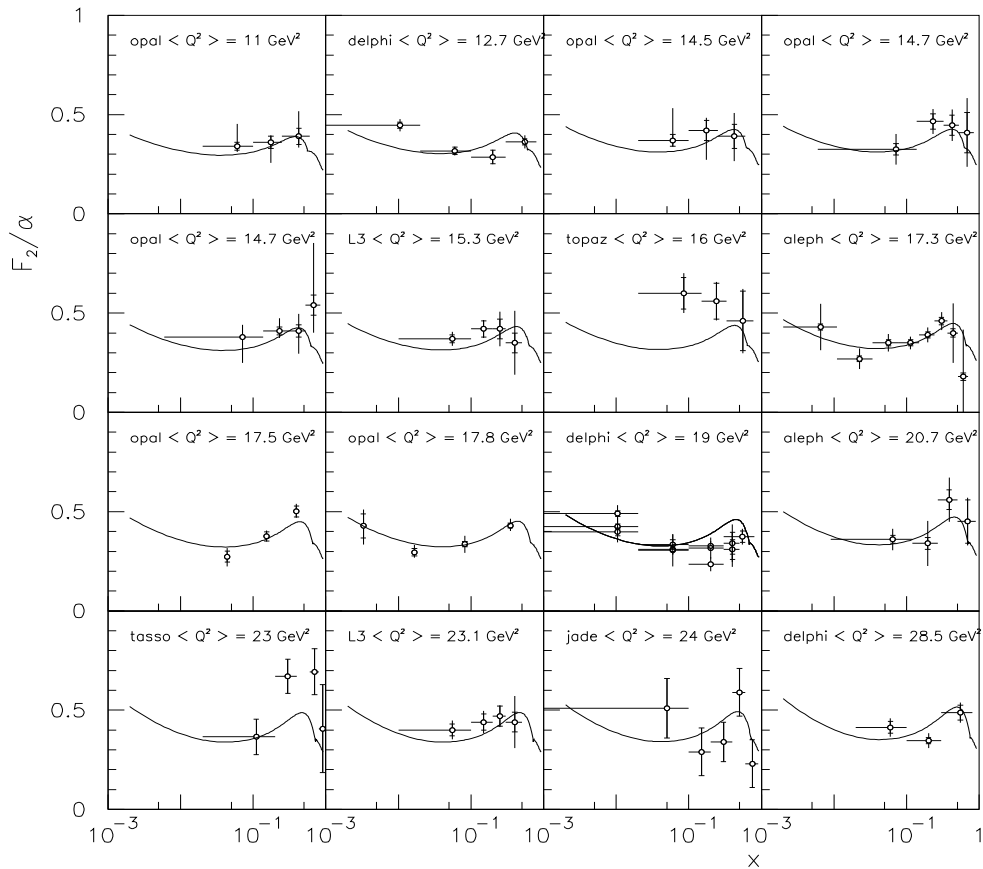}
\caption{Comparison between the best fit structure function and the
world data with 11 GeV$^2 < \ <\!Q^2\!>\  < 29$~GeV$^2$.}
\label{fig:8}
\end{figure}

\begin{figure}[htb]
\vspace{9pt}
\centering
\includegraphics[width=5.7in,height=5.2in]{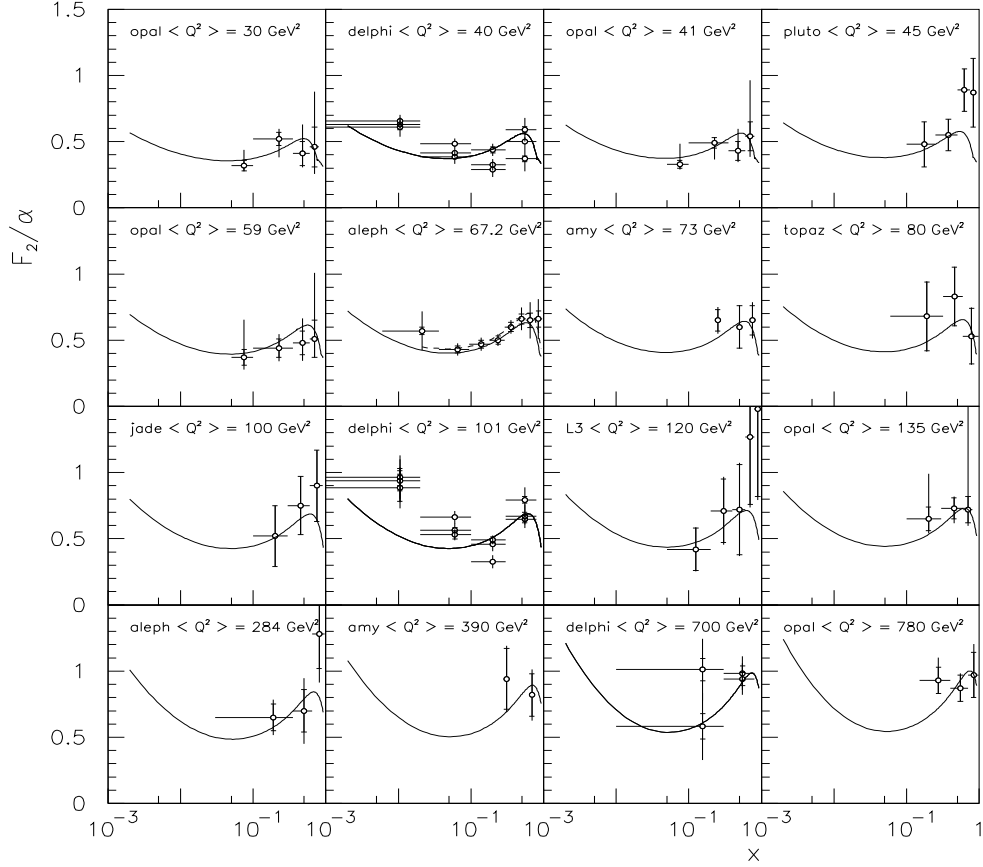}
\caption{Comparison between the best fit structure function and the
world data with 29 GeV$^2 < \ <\!Q^2\!>\  < 700$~GeV$^2$.}
\label{fig:9}
\end{figure}

\begin{figure}[htb]
\vspace{9pt}
\centering
\includegraphics[width=3in,height=2.5in]{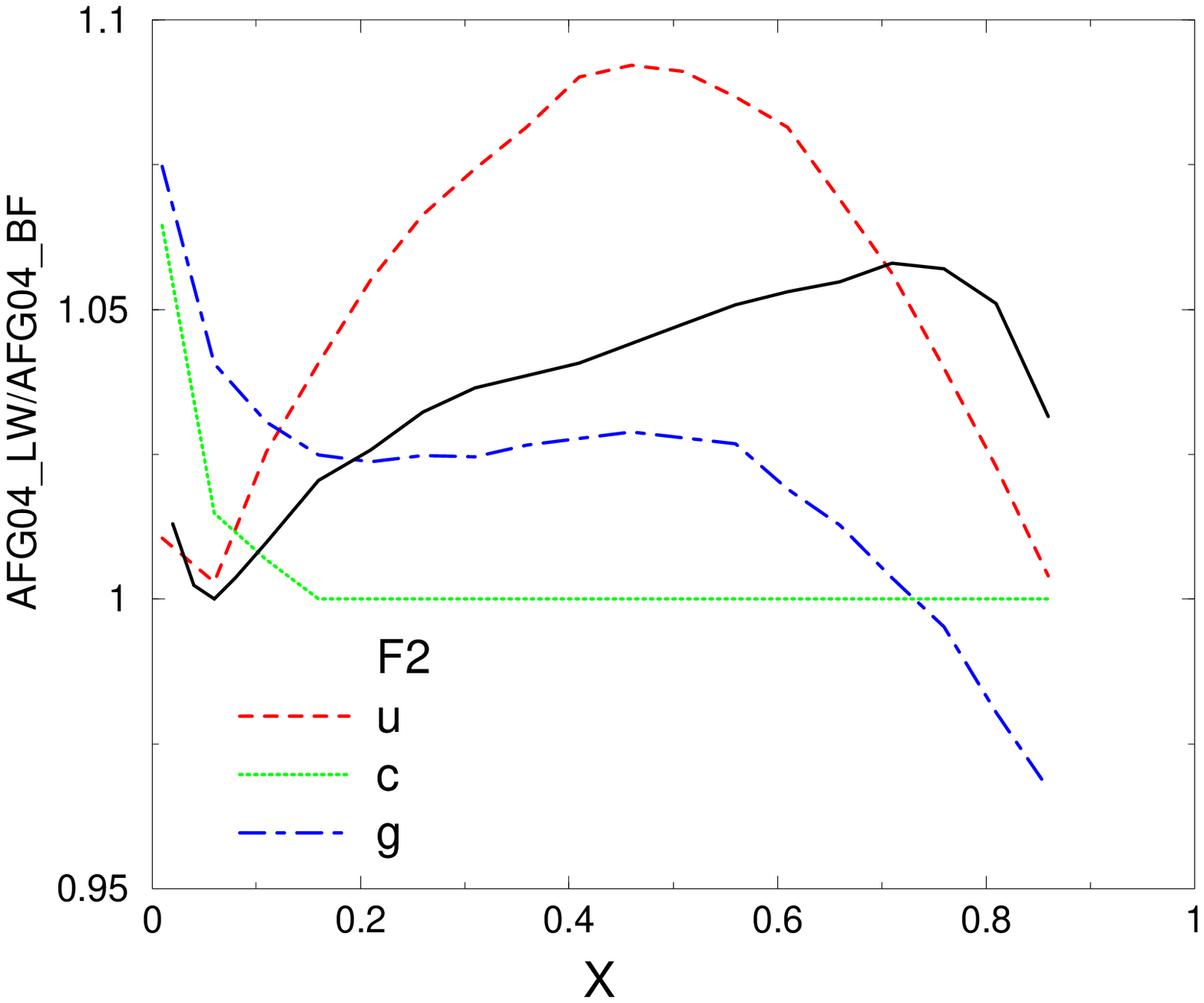}
\includegraphics[width=2.8in,height=2.5in]{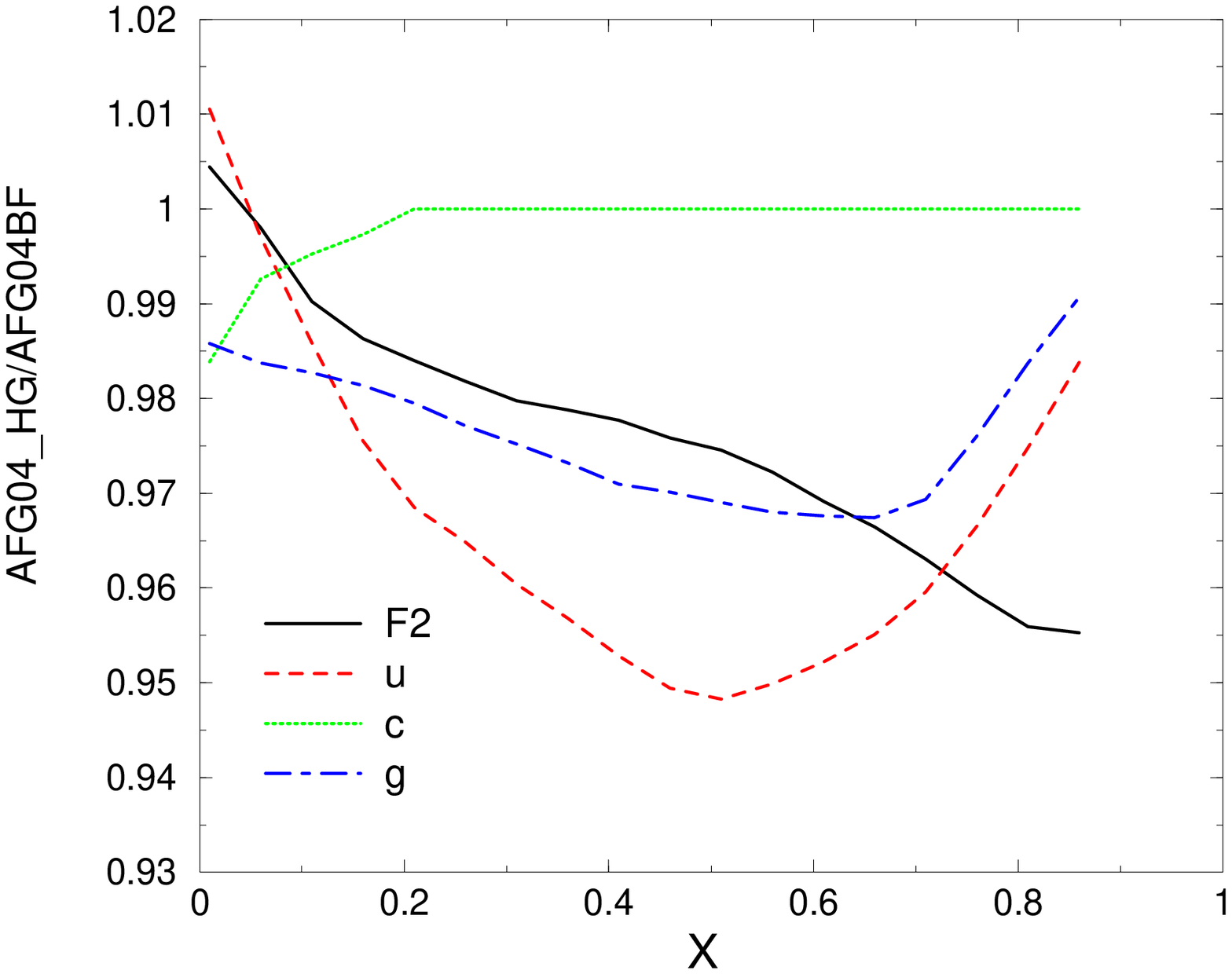}
\caption{The Low-$Q_0^2$ (left) and the High-$Q_0^2$ (right)
parametrizations compared to the best-fit parametrization.}\label{fig:10}
\end{figure}

Let us now compare, in Fig.~10, the Low-$Q_0^2$ and High-$Q_0^2$
AFG04 parametrizations to
the best-fit parametrization. Let us
consider the figure at the left. At small values of $x$ where the
non-perturbative component is large, the ratio partly reflects the
values of $C_{np}$ used in AFG04\_BF and AFG04\_LW. At very small values
of $x$, the perturbative contribution becomes
dominant and the ratio reflects the effect of the $Q^2$-evolution
which is larger for AFG04\_LW. At
$x \sim .5$, the non-perturbative contribution is smaller and the ratio
also reflects the effect of the $Q^2$-evolution of the perturbative
component. For large-$x$ values, the $u$-quark distribution contains a
term proportionnal to $\ln (1-x)$. Adding the contribution from
$k_q^{(1)}$ (\ref{2.13e}) to the $C_{\gamma ,c}^f$ contribution
(\ref{3.10e}), we obtain a term proportionnal to $\displaystyle{{\ln
n \over n}} (1 +
(\alpha_s(Q^2)/\alpha_s(Q_0^2))^{-2P_{qq}^{(0)}/\beta_0})$ which is
large for a small evolution $(P_{qq}^{(0)}(n)$ being negative at large $n$).
The $u$-quark ratio reflects this behavior. Finally we note that the
variations in the distribution functions never exceed 10~\%.

We end this section by a few words on other NLO parametrizations. Let
us start with the AFG one. In fact the AFG parametrization is very
close to the parametrization AFG04(.5, 1., 1.9). The only differences
come from the values of $\Lambda_{\overline{MS}}^{(4)}$ (300 MeV in
AFG04 and 200 MeV in AFG) and from the absence of bottom quark
distribution in AFG. The comparison is done at $Q^2 = 50$~GeV$^2$, a
scale which is in the range of those used in large-$p_{\bot}$
photoproduction. The ratio of the best-fit distribution AFG04\_BF to
the AFG distribution is displayed in Fig.~11. The smaller normalization
of the non-perturbative component of AFG04\_BF compared to the one used
in AFG explains the pattern at small-$x$ values. However, at very small
values of $x$, the inhomogeneous kernels $k_q^{(1)}$ and $k_g^{(1)}$
(\ref{2.4e} and \ref{2.5e}) have a singular behavior ($k_q^{(1)} \sim
\ln^2 x$ and $k_g^{(1)} \sim {1 \over x}$) and the behavior of the
perturbative gluon and sea components reflect a faster evolution of
AFG04 with $Q^2$ due to the larger value of $\Lambda_{\overline{MS}}$.
At large-$x$ values the perturbative contributions are important and
the ratio of the $u$-quark distribution is close to the ratio $r =
\alpha_s (AFG)/\alpha_s(AFG04)$ at $Q^2 = 50.$~GeV$^2$, namely $r \sim
.9$. The difference in the predictions for $F_2^{\gamma}$ are
illustrated in the figure on the right. It is not very large, and could
be distinguishable only at large values of $x$ where data are poor.
Comparisons between AFG04\_BF and AFG predictions are given for OPAL
(3.7), OPAL (10.7) and ALEPH (67.2) in Figs.~7 and 9. \par

A comparison with the GRS \cite{8r} parametrization can also be
performed on the basis of Figs.~7 and 8. In the second reference of
\cite{8r}, a comparison is made between OPAL data and the GRS predictions
which can hardly be distinguished from our best fit curves at $Q^2 =
3.7$, 10.7 and 17.8~GeV$^2$. However note that the large-$x$ behaviors
of $F_2^{\gamma}$ (outside the data range) are quite different between
the GRS and the AFG04\_BF parametrizations. This
behavior comes from the different factorization schemes used
($\overline{MS}$ versus $DIS_{\gamma}$ \cite{8r}), associated with
different non perturbative inputs (compare (\ref{3.11e}) and
(\ref{3.13e})).\par

In a recent publication, the authors of ref. \cite{reference11}
proposed new 5 flavor NLO parametrizations and did detailed
comparisons with world data, and other (AFG and GRS) parametrizations.
On the basis of Figs.~7, 8 and 9, and of similar figures in ref.
\cite{reference11}, we can easily observe several differences between
the parametrizations. At small $x$ ($x \ \lsim\ 10^{-2}$) and for $Q^2
\ \gsim \ 5$~GeV$^2$ the parametrizations of ref. \cite{reference11}
are higher than AFG04\_BF (which is very close to GRS). This trend
increases with $Q^2$. At medium $x$ in the charm threshold region and
at larger values of $x$ ($x \ \gsim\ .6$), the differences between
AFG04\_BF, and the FFNS$_{\rm CJK}1$ and CJK NLO parametrizations of
ref. \cite{reference11} are also noteworthy. Here also the origin of
this difference is the different non perturbative inputs associated
with the $\overline{MS}$ scheme (expression (\ref{3.11e})) and the
$DIS_{\gamma}$ scheme used in ref. \cite{reference11}. In all cases
data are not accurate enough to enable to distinguish between the
models.

\begin{figure}[htb]
\vspace{9pt}
\centering
\includegraphics[width=3in,height=2.5in]{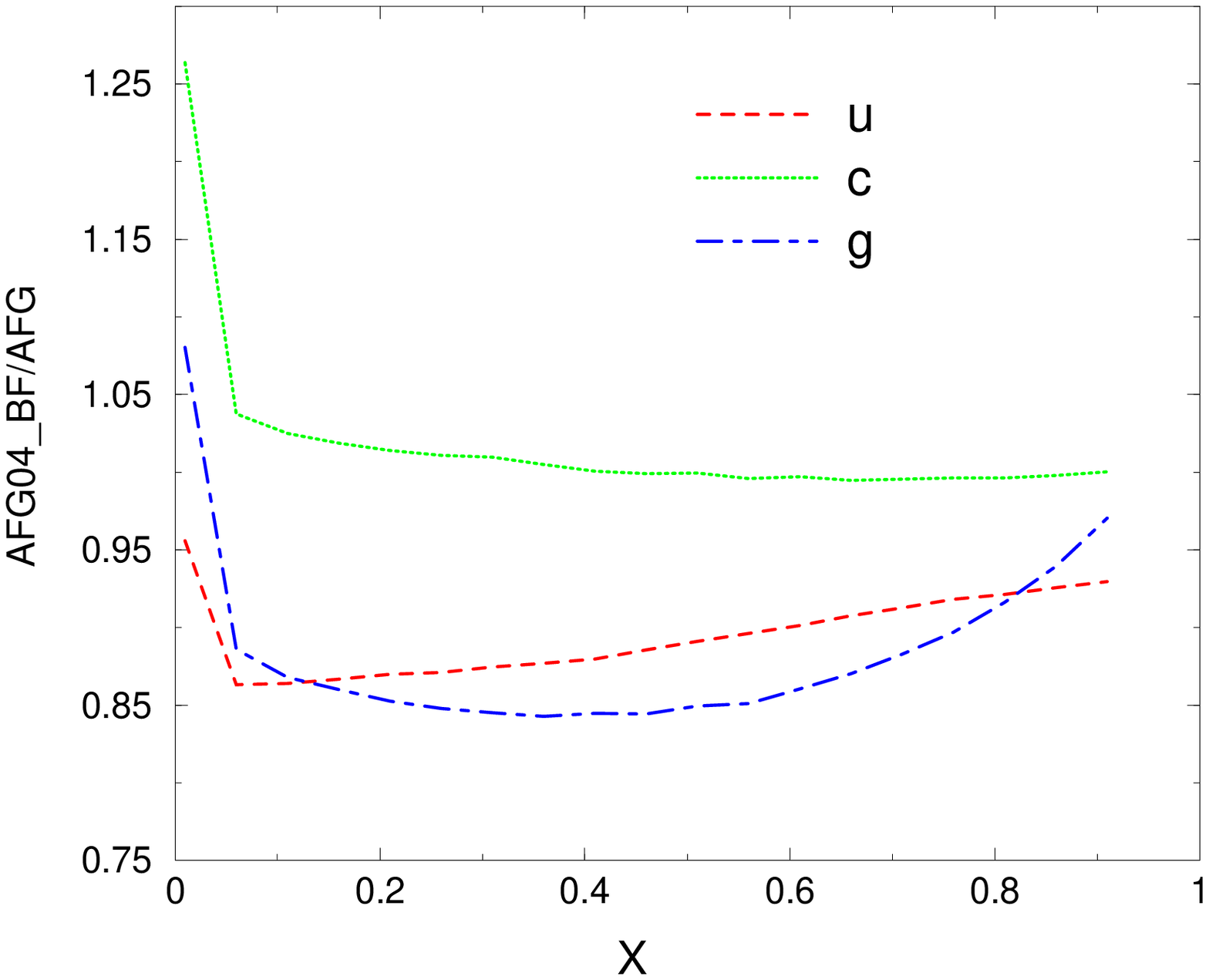}
\includegraphics[width=2.8in,height=2.5in]{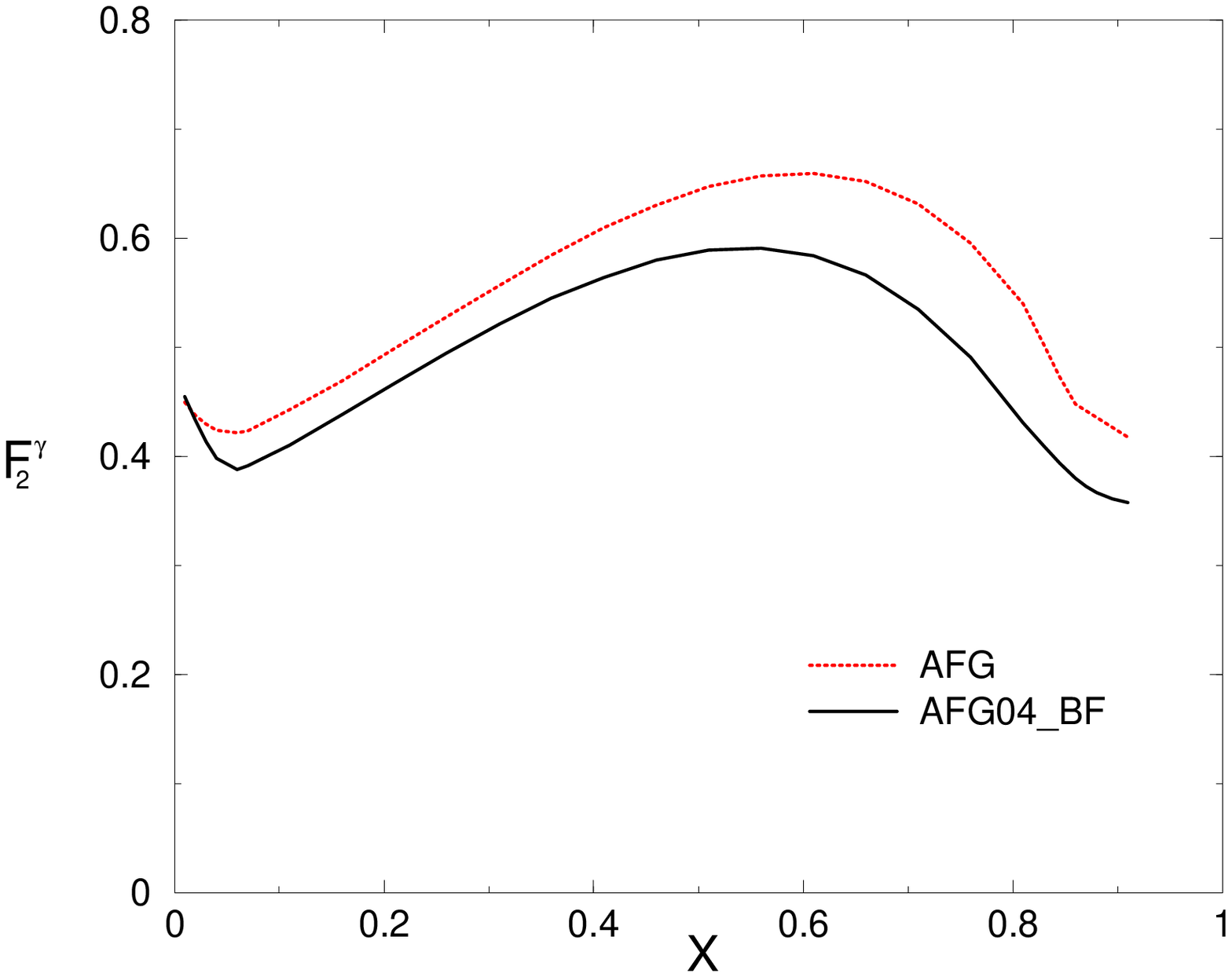}
\caption{The ratios AFG04\_BF/AFG at $Q^2 = 50.$~GeV$^2$(left) and
the structure functions $F_2^{\gamma} (x, Q^2)$ (right).}\label{fig:11}
\end{figure}

\mysection{The gluon content of the photon}
\hspace*{\parindent}
In this section we study other possible options for the parametrization
of the parton distributions in the photon. First we study a
modification of the sea quark distributions. Then we investigate the
effect of changing the normalization of the sea quark and gluon
distributions. And finally we modify the large-$x$ shape of the gluon
distribution. As we shall see, these modifications are poorly constrained
by $F_2^{\gamma}$ data, but some of them could be visible in
photoproduction experiments.\par

The small-$x$ behavior of the sea quark distribution (\ref{4.4be}) that
we used until now is less steep than those of some recent parton
distributions in the proton \cite{a,b,c}. For instance we have
$x(\overline{u} +
\overline{d}) \sim .061/x^{.3}$ at $x < 10^{-3}$ for the CTEQ6M
distribution at $Q_0^2 =
1.69$~GeV$^2$. In order to explore the effect of such a steep
behavior, we modify our sea distribution (\ref{4.4be}), while keeping
fixed the momentum carried by the sea quarks ($\int_0^1 dx \ x \
q_{sea}^{\pi} (x, Q_0^2 = 2$~GeV$^2) = .14$)

\beq
\label{6.1e}
xq_{sea}^{\pi} = .48 (1 - x)^{7.5} /x^{.3} \ .
\eeq

\noi This ansatz corresponds to quite a large sea at small-$x$, larger
by more than a factor 2 than the corresponding CTEQ6M parametrization
for the proton.
Therefore (\ref{6.1e}) must be considered as an extreme
parametrization.\par

The resulting $\chi^2$ is less satisfactory than the one obtained in the
preceding section. With the exception of DELPHI, the individual
$\chi^2$ deteriorate. The total best fit now corresponds to $\chi^2 =
34.3$ (without
DELPHI-errors modification), close to the value of
section 5. If we modify the errors, we obtain $\chi^2 = 31.4$,
instead of 25.8 in section 5. From these results, we see that there is
no compelling reason to modify the small-$x$ behavior of the sea
distributions we used in the preceding section. \par

\begin{figure}[htb]
\vspace{9pt}
\centering
\includegraphics[width=3in,height=2.in]{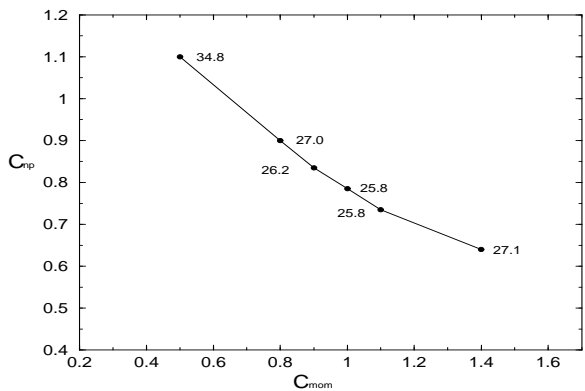}
\caption{Correlation between the parameters $C_{mom}$ and $C_{np}$.
The numbers are the $\chi^2$-values.}
\label{fig:12}
\end{figure}

Let us now study the effects of the parameter $C_{mom}$ which modifies
the normalization of the sea quark and gluon distributions according to
(\ref{4.7ae}, \ref{4.7be}) (at small-$x$ values, sea quark and gluon
distributions are strongly coupled and we modify their normalizations
by the same parameter $C_{mom}$). If we keep $Q_0^2$ and $C_{np}$ at
the best fit values established in section 5, $C_{mom}$ is quite
constrained by $F_2^{\gamma}$ data and the $\chi^2$-value varies by less
than one unit if we stay in the domain $.92 \ \lsim \ C_{mom} \
\lsim\ 1.08$. However it is clear that we can partially compensate the
$C_{mom}$ variations by also varying $C_{np}$. Keeping $Q_0^2 =
.7$~GeV$^2$, we first observe a strong correlation between $C_{np}$ and
$C_{mom}$ (Fig.~12). By playing with the values of $C_{mom}$ and
$C_{np}$, for instance, we can enhance the importance of the valence
compared to the sea quark. But it is unlikely that photoproduction
experiment could better constrain $C_{np}$ and $C_{mom}$ and we do not
pursue this study in detail.\par

\begin{figure}[htb]
\vspace{9pt}
\centering
\includegraphics[width=3in,height=2.in]{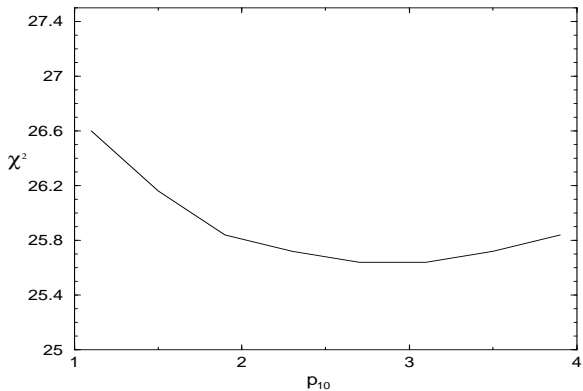}
\caption{$\chi^2$ variation as a function of the gluon parameter
$p_{10}$.} \label{fig:13}
\end{figure}

Finally we consider the modification of the gluon distribution and we
vary the parameter $p_{10}$ of expression
(\ref{4.4ce}). As expected $F_2^{\gamma}$ is not sensitive to the gluon
distribution and LEP data do not constrain the value of $p_{10}$. We
display in Fig.~13 the dependence of the $\chi^2$ on $p_{10}$ which
is weak ($Q_0^2$ and $C_{np}$
being fixed at the best fit values). In a
large range in $p_{10}$, $\chi^2$ varies by less than one unit. In
Fig.~14 we show the behaviour of the distributions obtained with
$p_{10} = 1.0$ (hard gluon) and $p_{10} = 4.0$ (soft gluon) at $Q^2 =
50.$~GeV$^2$, the other parameters being kept fixed at the best-fit
values. The behavior at small values of $x$ is due to the normalization
factor $C_g$ (\ref{4.4ce})~; at large values of $x$ the
non-perturbative inputs vanish and the ratios go to one.

\begin{figure}[htb]
\vspace{9pt}
\centering
\includegraphics[width=2.8in,height=2.8in]{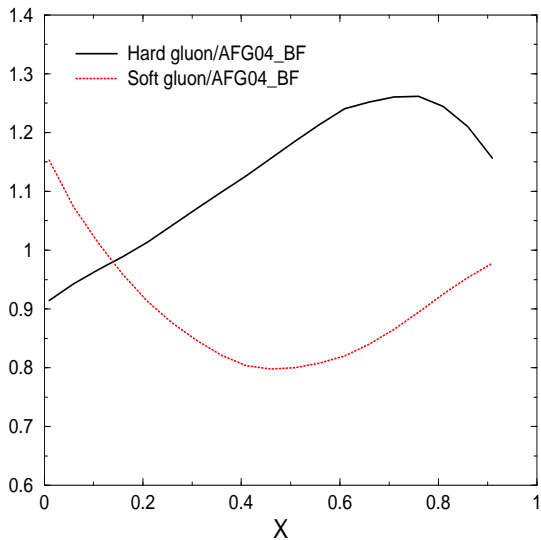}
\caption{Ratios of the gluon distributions with $p_{10} = 4.0$ (soft
gluon) and $p_{10} = 1.0$ (hard gluon) to the best-fit gluon at $Q^2
= 50.$~GeV$^2$.} \label{fig:14}
\end{figure}

On the other hand photoproduction reactions \cite{new50} are sensitive
to the gluon distribution since an initial gluon can interact with a
parton from the initial proton producing two large-$p_{\bot}$ jets in 
the final state. Jet production in photon-photon
collisions is also a reaction allowing to observe the gluon
distribution in the range $0 < x\ \lsim\ .4$ \cite{new51}. A
particularly interesting reaction is the
photoproduction of large-$p_{\bot}$ photon and jet, which has been
studied in ref. \cite{9r} in detail. The interest of this reaction
comes from the fact that the scale dependence of the cross section is
well under control, and therefore, the theoretical predictions are
reliable. We quote here one result of this paper, referring the
interested reader to the original publication \cite{9r}. Fig.~15
displays the cross sections $d\sigma/dx_{LL}$, where $x_{LL} =
p_{\bot}^{\gamma}$ ($e^{-\eta_{\gamma}} +
e^{-\eta_{jet}})/2E_{\gamma}$, for various cuts on the rapidities
$\eta^{\gamma}$ and $\eta^{jet}$ ($E_{\gamma}$ is the energy of the
initial photon). In the forward region where the rapidities
$\eta^{\gamma}$, $\eta^{jet}$ are large, the cross section is dominated
by the resolved contribution. For some cuts, half of the cross section
is due to the gluon distribution in the photon. But the observable
range in $x$ is small ($x \ \lsim\ .2$) and the cross sections fairly
small. However this type of data could be used to constrain the poorly
known $G^{\gamma}(x,Q^2)$.

\begin{figure}[htb]
\vspace{9pt}
\centering
\includegraphics[width=4in,height=4in]{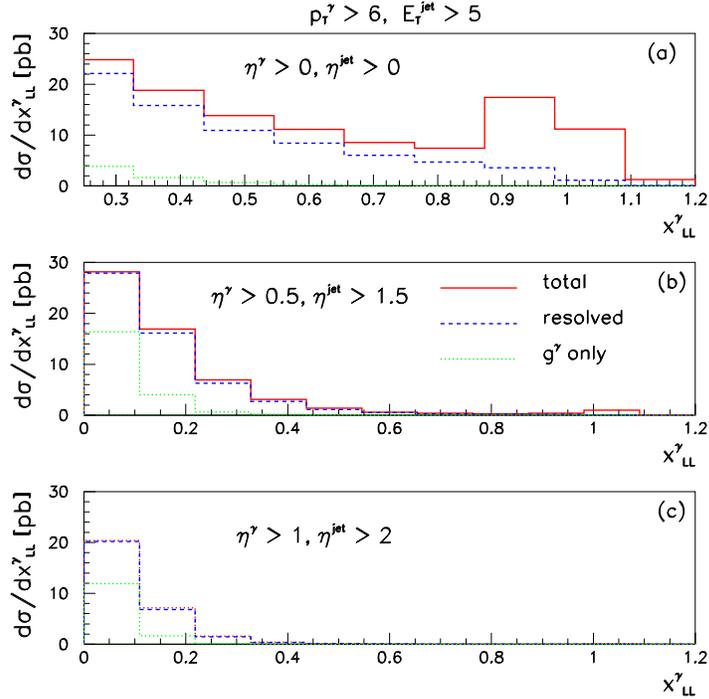}
\caption{The enhancement of the gluon contribution to the reaction
$\gamma + p \to \gamma + jet + X$ due to cuts on the rapidities
$\eta_{\gamma}$ and $\eta_{jet}$.} \label{fig:15}
\end{figure}
\section{Conclusions}

We have proposed a new set of next-to-leading order parton
distributions for real photons. We work in the $\overline {MS}$ scheme
with a variable number of massless flavors, keeping however the massive
correction terms for charm and beauty quarks, in the direct
contribution to the ${\cal{F}}_2^\gamma$ structure function. The
perturbative contribution is assumed to vanish below a $Q_0^2$ value
and a VDM type parametrisation is used for the non-perturbative input
at this value. We give a detailed discussion on how to isolate the
scheme invariant piece of the structure function to which the physical
VDM input should be applied. The distributions are set up such that
various input parameters can be easily changed: these are the value of
$Q_0^2$, the overall normalisation of the non-perturbative input and
the x-dependence of the gluon which is poorly constrained by the photon
structure function data. Using the LEP data, an attempt to decorrelate
the perturbative contribution, controlled by $Q_0^2$, from the
non-perturbative one is made. It turns out that the error bars are too
large to determine unambiguously the value of $Q_0^2$. A best fit
to low and medium-$Q^2$ LEP data is then done which is also shown to 
yield a good
agreement with the world data. A large dispersion in the value of the
best fit parameters is observed when a minimum $\chi^2$ fit is
performed for each set of LEP data independently. We therefore propose
several sets of parton distributions to ``frame'' the data. Since the
gluon distribution cannot be constrained from the deep inelastic  data,
we suggest to consider the photoproduction of photons, hadrons or jets
at large transverse momentum.

\newpage
\begin{appendix}
\mysection{- Appendix A}
\hspace*{\parindent} In this appendix, we give a derivation of
expression (\ref{3.10e}). We
start from the reaction in which the target photon, instead of being
real, has a small virtuality $p^2$ $(-p^2 \equiv P^2 \ll Q^2)$, but
large enough for the perturbative approach to be valid. This allows us
to study the structure of the HO corrections to ${\cal F}_2^{\gamma}$
and to understand how to take the real photon limit $p^2 \to 0$. The
transverse structure function ${\cal F}_2^T(Q^2, P^2, n)$ (transverse
with respect to the polarisation of the target photon) can be written

\beq
\label{A.1e}
{\cal F}_2^T(Q^2, P^2,n) = \sum_f e_f^2\ C_q(n)\ q_f^{NS}(Q^2,P^2, n)
+ C_{\gamma}^{NS}(n)
\eeq

\noi where, for simplicity, we only consider the Non Singlet contribution. \par

This expression is obtained in resumming all the $\ln (Q^2/P^2)$ in the
quark distribution $q_f^{NS}(Q^2, P^2, n)$, whereas $C_q(n)$ and
$C_{\gamma}^{NS}(n)$ are expansions in power of $\alpha_s(Q^2)$. This
procedure defines a factorization scheme called ``Virtual Factorization
Scheme'' in ref. \cite{A1}. In this scheme the Wilson coefficient $C_q(n)$ and
the direct term $C_{\gamma}^{NS}(n)$ are known since the work of
Uematsu and Walsh \cite{A2}. \par

The quark distribution is a solution of eq. (\ref{2.2e}) (we drop the
index NS and the moment variable $n$)

\beq
\label{A.2e}
q_f(Q^2,P^2) = \sigma_f\int_{\alpha_s(P^2)}^{\alpha_s(Q^2)} {d \lambda \over
\beta (\lambda )} \ k_{q} (\lambda ) \
e^{\int_{\lambda}^{\alpha_s(Q^2)} {d\lambda ' \over \beta (\lambda
')}\ P_{qq}(\lambda )} \ .
\eeq

To simplify the notation further, we consider one quark species and drop
the charge factors which are present in $\sigma_f k_q(\lambda )$ of
(\ref{A.2e}) (that we
shall note $k(\lambda )$) and in $C_{\gamma}^{NS}$. With
this convention, the direct term is written
\cite{A2}\bea\label{A.3e}C_{\gamma}(x) &
=& {\alpha \over 2 \pi}\ 6 \Big \{ \left [ (x^2 + (1-
x)^2) \ln {1 \over x} + 2x(1-x)-1\right ] \nn \\
&&+ \left ( x^2 + (1-x)^2\right ) \ln {1 \over x} + 6x(1-x)-1 \Big \} \ .
\eea

It is worth noting the following points. First $C_{\gamma}(x)$ is
scheme dependent and different from the $\overline{\rm MS}$ expression
(\ref{2.8e}). However it is easy to move from the Virtual Scheme to the
$\overline{\rm MS}$ scheme by keeping in mind that expression (\ref{2.15e})
must be scheme invariant, which leads to

\beq
\label{A.4e}
{\alpha \over 2 \pi} \left ( {k^{(1)^{\overline{MS}}}(n)
\over P_{qq}^{(0)}(n)} - {k^{(1)}(n) \over
P_{qq}^{(0)}(n)}\right ) = C_{\gamma}^{\overline{MS}}(n) -
C_{\gamma}(n) \equiv \Delta C(n)
\eeq

\noi where, from (\ref{2.8e}) and (A.3)

\beq
\label{A.5e}
\Delta C(x) = {\alpha \over 2 \pi} 6 \left \{ \left ( x^2 + (1 -
x)^2\right ) \ln\ x(1-x) + 1 \right \}
\eeq

\noi in the $x$-space.\par

Second, the direct terms $C_{\gamma}$ are target dependent and depend on
the regularization used to avoid a collinear divergence in the
calculation of the box diagram. For instance, in dimensional
regularization, the target dependent part has been calculated in section
3, and is given by $C_{\gamma , c}^f$ (\ref{3.7e}) (in which $e_f^4$
is dropped). For a virtual
photon, we obtain the first line of (\ref{A.3e}) \cite{A1}, whereas the second
line is universal and equal to the equivalent $\overline{\rm MS}$
expression. Therefore we can write in general

\beq
\label{A.6e}
C_{\gamma}(x) = C_{\gamma}^{col}(x) + C_{\gamma}^U(x) \ .
\eeq

\noi With the notation ``col'', we indicate that the target dependent
part of $C_{\gamma}$ comes from the lower limit of the integration over
$k^2$ (cf. expression \ref{3.2e} and \ref{3.6e}). A similar
decomposition exists for $k^{(1)}$

\beq
\label{A.7e}
k^{(1)} = k^{(1)^{col}} + k^{(1)^U}
\eeq

\noi because the target dependent terms, which are present in
$C_{\gamma}(x)$, also appear in the course of the calculation of
$k^{(1)}$ under the form $C_{\gamma}^{col}(n)$
$P_{qq}^{(0)}(n)$ (this point has been discussed in detail in ref.
\cite{A1}). Therefore the combination

\beq
\label{A.8e}
- {\alpha \over 2 \pi}\ {k^{(1)}(n) \over P_{qq}^{(0)}(n)} +
C_{\gamma}(n) = - {\alpha \over 2 \pi} \ {k^{(1)^U}(n) \over
P_{qq}^{(0)}(n)} + C_{\gamma}^U(n)
\eeq

\noi and consequently ${\cal F}_2(Q^2, P^2,n)$ (when $Q^2$ is very
large) does not depend on the details of the target. \par

We are now
ready to study the limit $P^2 \to 0$ of expression (\ref{A.2e}). With
this aim in view, we introduce an intermediate scale $Q_0^2$ which
allows us to isolate the $P^2$-dependent part of (\ref{A.2e})

\beq
\label{A.9e}
q(Q^2, P^2) = q(Q^2, Q_0^2) +
\int_{\alpha_s(P^2)}^{\alpha_s(Q_0^2)}{d\lambda \over \beta
(\lambda)} \ k({\lambda}) \ e^{\int_{\lambda}^{\alpha_s(Q^2)}
{d \lambda ' \over \beta (\lambda ')} P_{qq}(\lambda ')} \ ,
\eeq

\noi where $q(Q^2, Q_{0}^2)$ is given by (\ref{A.2e}) in which $P^2$
is replaced by $Q_0^2$ (in section 3, $Q_0^2$ is the limit between
the perturbative and the non-perturbative domains). Using the notation

\beq
\label{A.10e}
k(\lambda ) = {\alpha \over 2 \pi} \left \{
k^{(0)} + {\lambda \over 2 \pi} \ k^{(1)^U} +
{\lambda \over 2 \pi}\ k^{(1)^{col}}\right \} =
k^U(\lambda ) + {\alpha \over 2 \pi} \ {\lambda \over 2
\pi}\ k^{(1)^{col}}
\eeq
\noi we rewrite the integral in (A.9)

\beq
\label{A.11e}
\left \{ \int_{P^2}^{Q_0^2} {dk^2 \over k^2} k^U (\lambda )
\ e^{\int_{k^2}^{Q_0^2}{dk'^2 \over k'^2} P_{qq}} - {\alpha \over 2
\pi}\ {k^{(1)^{col}} \over P_{qq}^{(0)}} \left ( 1 - \left
( {\alpha_s(Q_0^2) \over \alpha_s(P^2)}\right )^{-2 {P_{qq}^{(0)}
\over \beta_0}}\right ) \right \}\cdot e^{\int_{\alpha
(Q_0^2)}^{\alpha (Q^2)} {d \lambda ' \over \beta (\lambda ')}\ P_{qq}(\lambda
')}\ .
\eeq

After extracting from (\ref{A.11e}) the $P^2$-independent, but target
dependent term
$k^{(1)^{col}}/P_{qq}^{(0)}$, we define

\beq
\label{A.12e}
H(Q_0^2, P^2) = \int_{P^2}^{Q_0^2} {dk^2 \over k^2} \ k^U \
e^{\int_{k^2}^{Q_0^2} {dk'^2 \over k'^2}P_{qq}} + {\alpha \over 2
\pi}\ {k^{(1)^{col}} \over P_{qq}^{(0)}}  \left (
{\alpha_s(Q_0^2) \over \alpha_s(P^2)}\right )^{- 2 {P_{qq}^{(0)}
\over \beta_0}}
\eeq

\noi and rewrite ${\cal F}^T_2$ as

\beq
\label{A.13e}
{\cal F}_2^T(Q^2, P^2,n) = C_q \ q(Q^2, Q_0^2) + C_q \left [ H(Q_0,
P^2) - C_{\gamma}^{col} \right ]
e^{\int_{\alpha_s(Q_0^2)}^{\alpha_s(Q^2)} {d \lambda \over \beta
(\lambda )} P_{qq}} + C_{\gamma}^U + C_{\gamma}^{col} \ ,
\eeq

\noi because $C_{\gamma}^{col} = {\alpha \over 2\pi} {k^{(1)^{col}}
\over P_{qq}^{(0)}}$. Let us now consider the limit $P^2 \to 0$
with $Q_0^2$ being the scale below which the perturbative approach has
no meaning. As a consequence, the perturbative expression (\ref{A.12e})
of $H(Q_0^2, P^2)$ is no longer valid and we can only say that $H(Q_0^2,
P^2)$ contains all the non-perturbative contributions needed to define
${\cal F}_2^{\gamma}$ in the real limit. Now we recognize in
(\ref{A.13e}) the structure of the input proposed in formula
(\ref{3.10e}) if we identify $H(Q_0^2) = \displaystyle{\lim_{P^2 \to
0}} H(Q_0^2, P^2)$ to $q^{NP}(Q_0^2) + \overline{q}^{NP}(Q_0^2)$. \par

Expression (\ref{A.13e}) has been established in the virtual
factorization scheme. But we can easily obtain a similar expression in
the $\overline{\rm MS}$ scheme by writing $k^{(1)} =
k^{(1)^{\overline{MS}}} + \Delta k^{(1)}$ in the expression
(\ref{A.2e}) for $q (Q^2, Q_0^2)$. This change generates a
$\overline{\rm MS}$ distribution $q^{\overline{MS}}(Q^2, Q_0^2)$ and a
term (cf. (\ref{2.13e}))

\beq
\label{A.16e}
- {\Delta k^{(1)} \over P_{qq}^{(0)}} \left ( 1 - \left ( {\alpha_s(Q^2) \over
\alpha_s(Q_0^2)}\right )^{-2P_{qq}^{(0)}/\beta_0}\right )
\eeq

\noi which is combined with $C_{\gamma}^{col}\left (1 - \left (
{\alpha_s(Q^2) \over \alpha_s(Q_0^2)}\right
)^{-2P_{qq}^{(0)}/\beta_0}\right )$ of (\ref{A.13e}) to give

$$C_{\gamma}^{col, \overline{MS}}\left ( 1 - \left ( {\alpha_s(Q^2)
\over \alpha_s(Q_0^2)}\right )^{-2 P_{qq}^{(0)}/\beta_0}\right ) $$

\noi so that (\ref{A.13e}) can be written in terms of $\overline{MS}$
perturbative expressions, together with a non-perturbative contribution
$H(Q_0^2)$ which remains invariant under this change.

\newpage

\mysection{- Appendix B}
\hspace*{\parindent}
The parametrizations discussed in this paper are available in the form
of a FORTRAN code allowing the users to select the parton distributions
they are interested in. The conventions we use are described in table~1
which displays abbreviations of the general notation AFG04($Q_0^2,
C_{np}, p_{10}$).

\vskip 1 truecm
\begin{center}
\begin{tabular}{|c|c|c|c|}
\hline
&&&\\
&$Q_0^2 = .34$~GeV$^2$ &$Q_0^2 = .70$~GeV$^2$ &$Q_0^2 = .97$~GeV$^2$\\
&&& \\
\hline
$p_{10} = 1.0$ & &AFG04\_BF\_1.0 &\\
(hard gluon) & &($C_{np} = .78$) & \\
\hline
$p_{10} = 1.9$ &AFG04\_LW &AFG04\_BF &AFG04\_HG\\
(default) &($C_{np} = .60$) &($C_{np} = .78$) &($C_{np} = .84$)\\
\hline
$p_{10} = 4.0$ & &AFG04\_BF\_4.0 &\\
(soft gluon) & &($C_{np} = .78$) &\\
\hline
\end{tabular}
\par\vskip 5 truemm
{Table 1}
\end{center}
\vskip 5 truemm

\noi The values of $C_{np}$ indicated in brackets are the default
values~; the users can choose other values (after carefully reading
section 5). These parametrizations can be downloaded from the site
http://www.lapp.in2p3.fr/lapth/PHOX\_FAMILY/main.html.

\end{appendix}


\newpage

\begin{thebibliography}{99}

\bibitem{2r} E. Witten, Nucl. Phys. {\bf B120}, 189 (1977).

\bibitem{1r} H. Kolanoski, Springer Tract in Modern Physics {\bf
105}, 187 (1984)~; \\
Ch. Berger and W. Wagner, Phys. Rep.  {\bf C146}, 1 (1987).
 
 
\bibitem{3r} R. Nisius, Phys. Rep. {\bf 332}, 165 (2000).

\bibitem{4r} M. Krawczyk, M. Staszel and A. Zembrzuski, Phys. Rep.
{\bf 345}, 265 (2001).

\bibitem{5r} M. Klasen, Rev. Mod. Phys. {\bf 74}, 1221 (2002).

\bibitem{6r} P. Aurenche, M. Fontannaz, J. Ph. Guillet, Z. Phys. {\bf
C64}, 621 (1994).

\bibitem{9r} M. Fontannaz and G. Heinrich, Eur. Phys. J. {\bf C34},
191 (2004).

\bibitem{10r} H1 Collaboration, C. Adloff et al., hep-ex/0302034.

\bibitem{11r} P. Aurenche et al., Phys. Lett. {\bf B233}, 517 (1989).

\bibitem{8r} M. Gl\"uck, E. Reya and I. Schienbein, Phys. Rev. {\bf
D60}, 054019 (1999), Erratum-ibid. {\bf D62}, 019902 (2000)~; Phys.
Rev. {\bf D64}, 01750 (2001).

\bibitem{reference11} F. Cornet, P. Jankowski, M. Krawczyk, Phys.
Rev. {\bf D70}, 093004 (2004).

\bibitem{new10} R. De Witt et al., Phys. Rev. {\bf D19}, 2046
(1979)~; (E) {\bf 20}, 1751 (1979).

\bibitem{new11} E. G. Floratos, D. A. Ross, C. T. Sachrajda, Nucl.
Phys. {\bf B129}, 66 (1977)~; E {\bf B139}, 545 (1978)~; Nucl. Phys.
{\bf B152}, 493 (1979).

\bibitem{new12} G. Curci, W. Furmanski, R. Petronzio, Nucl. Phys.
{\bf B175}, 27 (1980).\\
W. Furmanski, R. Petronzio, Phys. Lett. {\bf 97B}, 437 (1980).

\bibitem{12r} M. Fontannaz and E. Pilon, Phys. Rev. {\bf D45}, 382 (1992).

\bibitem{13r} M. Gl\"uck, E. Reya and A. Vogt, Phys. Rev. {\bf D45},
3986 (1992).

\bibitem{new15} NLO FORTRAN codes for large-$p_T$ photoproduction
reactions can be downloaded from
http://www.lapp.in2p3.fr/lapth/PHOX\_FAMILY/main.html.

\bibitem{new16} E. B. Zijlstra and W. L. van Neerven, Nucl. Phys.
{\bf B383}, 525 (1992).

\bibitem{new17} S. Moch, J. A. M. Vermaseren, A. Vogt, Nucl. Phys.
{\bf B621}, 413 (2002).

\bibitem{new18} S. Moch, J. A. M. Vermaseren, A. Vogt, Nucl. Phys.
{\bf B688}, 101 (2004).\\
A. Vogt, S. Moch, J. A. M. Vermaseren, Nucl. Phys. {\bf B691}, 129 (2004).

\bibitem{14r} W. A. Bardeen, A. J. Buras, D. W. Duke and T. Muta,
Phys. Rev. {\bf D18}, 3998 (1978).\\
G. Altarelli, R. K. Ellis and G. Martinelli, Nucl. Phys. {\bf B157},
461 (1979).

\bibitem{new.a} W. A. Bardeen, A. J. Buras, Phys. Rev. {\bf D20}, 166
(1979)~; {\bf D21}, 2041 (1980)E.

\bibitem{7r} M. Gl\"uck, E. Reya and A. Vogt, Phys. Rev. {\bf D46},
1973 (1992).

\bibitem{new23} E. Witten, Nucl. Phys. {\bf B104}, 445 (1976)~; \\
   C. T. Hill and G. C. Ross, Nucl. Phys. {\bf B148}, 373 (1979)~;\\
M. Gl\"uck and E. Reya, Phys. Lett. {\bf B83}, 98 (1979).

\bibitem{Z} S. Albino, M. Klasen and S. S\"oldner-Rembold, Phys.
Rev. Lett. {\bf 89}, 122004 (2002).

\bibitem{16r} ALEPH Collaboration, A. Heister et al., Eur. Phys. J.
{\bf C30}, 145 (2003).

\bibitem{17r} DELPHI Collaboration, preprint DELPHI 2003-025 CONF
645, contributed paper for EPS 2003 (Aachen) and LP 2003 (FNAL).


\bibitem{18r} L3 Collaboration, M. Acciarri et al., Phys. Lett. {\bf
B447}, 147 (1999).

\bibitem{19r} OPAL Collaboration, G. Abbienchi et al., Eur. Phys. J.
{\bf C18}, 15 (2000).

\bibitem{20r} PLUTO Collaboration, Ch. Berger et al., Phys. Lett.
{\bf B142}, 111 (1984).

\bibitem{21r} L3 Collaboration, M. Acciarri et al., Phys. Lett. {\bf
B436}, 403 (1998).

\bibitem{22r} TPC-2$\gamma$ Collaboration, H. Aihara et al., Z. Phys.
{\bf C34}, 1 (1987).

\bibitem{22NEW} L3 Collaboration, M. Acciarri et al., Phys. Lett.
{\bf B483}, 373 (2000).

\bibitem{23NEW} AMY Collaboration, Sasaki et al., Phys. Lett. {\bf
252B}, 491 (1990).

\bibitem{24NEW} AMY Collaboration, Sahu et al., Phys. Lett. {\bf
346B}, 208 (1995).

\bibitem{25NEW} AMY Collaboration, Kojima et al., Phys. Lett. {\bf
B400}, 395 (1997).

\bibitem{26NEW} DELPHI Collaboration, Abreu et al., Zeit. Phys. {\bf
C69}, 223 (1996).


\bibitem{27NEW} JADE Collaboration, Bartel et al., Zeit. Phys. {\bf
C24}, 231 (1984)~; Phys. Lett. {\bf B121}, 203 (1983).


\bibitem{28NEW} OPAL Collaboration, Akers et al., Zeit. Phys. {\bf
C61}, 199 (1994).

\bibitem{29NEW} OPAL Collaboration, Ackerstaff et al., Zeit. Phys.
{\bf C74}, 33 (1997).

\bibitem{30NEW} OPAL Collaboration, Ackerstaff et al., Phys. Lett.
{\bf B411}, 387 (1997).

\bibitem{31NEW} OPAL Collaboration, Ackerstaff et al., Phys. Lett.
{\bf B412}, 225 (1997).

\bibitem{32NEW} OPAL Collaboration, Abbiendi et al., Phys. Lett.
{\bf B533}, 207 (2002).

\bibitem{33NEW} PLUTO Collaboration, Ch. Berger et al., Nucl. Phys.
{\bf B281}, 365 (1987).

\bibitem{34NEW} TASSO Collaboration, Althoff et al., Zeit. Phys. {\bf
C31}, 527 (1986).

\bibitem{35NEW} TOPAZ Collaboration, Muramatsu et al., Phys. Lett.
{\bf 332B}, 477 (1994).

\bibitem{new49} M. Fontannaz, J. Ph. Guillet, G. Heinrich, Eur. Phys.
J {\bf C26}, 209 (2002).

\bibitem{a} V. Barone, C. Pascaud, F. Zomer, Eur. Phys. J {\bf C12},
243 (2000).

\bibitem{b} CTEQ Collaboration, J. Pumplin et al., JHEP {\bf 0207}, 012 (2002).

\bibitem{c} MRST Collaboration, A. D. Martin et al., Eur. Phys. J
{\bf C23}, 73 (2002).



\bibitem{new50} P. Aurenche, L. Bourhis, M. Fontannaz, J. Ph.
Guillet, Eur. Phys. J {\bf C17}, 413 (2000).

\bibitem{new51} OPAL Collaboration, G. Abbiendi et al., Eur. Phys. J.
{\bf C31}, 307 (2003).

\bibitem{A1} M. Fontannaz,  Eur. Phys. J. {\bf C38}, 297 (2004).

\bibitem{A2} T. Uematsu and T. F. Walsh, Nucl. Phys. {\bf B199}, 93 (1982).


\end{thebibliography}
\end{document}